\documentclass{aa}  

\usepackage{graphicx}
\usepackage{txfonts}
\usepackage{epsfig}
\usepackage{color}
\usepackage{amsmath}
\usepackage{amssymb}
\usepackage{natbib}

\def\lsim{\mathrel{\rlap{\lower3.5pt\hbox{\hskip0.5pt$\sim$}}
    \raise0.5pt\hbox{$<$}}}
\def\gsim{~\rlap{$>$}{\lower 1.0ex\hbox{$\sim$}}}

\def\ls{\mathrel{\lower0.6ex\hbox{$\buildrel {\textstyle <}\over{\scriptstyle \sim}$}}}
\def\gs{\mathrel{\lower0.6ex\hbox{$\buildrel {\textstyle >}\over{\scriptstyle \sim}$}}}

\begin{document}

\title{Mass\,-\,concentration relation and weak lensing peak counts}

\author{V.F. Cardone \inst{1} \thanks{Corresponding author\,: email {\tt winnyenodrac@gmail.com}}
\and S. Camera \inst{2}
\and M. Sereno \inst{3}
\and G. Covone \inst{4,5}
\and R. Maoli \inst{6}
\and R. Scaramella \inst{1}}

\institute{I.N.A.F.\,-\,Osservatorio Astronomico di Roma, via Frascati 33, 00040 Monte Porzio Catone (Roma), Italy
\and CENTRA, Instituto Superior T\'ecnico, Universidade de Lisboa, Av. Rovisco Pais 1, 1049-001 Lisboa, Portugal
\and Dipartimento di Fisica e Astronomia, Universit\`a di Bologna, Viale Berti Pichat 6/2, 40127 Bologna, Italy
\and Dipartimento di Fisica, Universit\`{a} di Napoli ``Federico II'', Compl. Univers. di Monte S. Angelo, via Cinthia, 80126 - Napoli, Italy
\and Istituto Nazionale di Fisica Nucleare, Sezione di Napoli, Compl. Univers. di Monte S. Angelo, via Cinthia, 80126 Napoli, Italy
\and Dipartimento di Fisica, Universit\`{a} di Roma ``La Sapienza'', Piazzale Aldo Moro, 00185 Roma, Italy }

\date{Accepted xxx, Received yyy, in original form zzz}

\abstract
{The statistics of peaks in weak lensing convergence maps is a promising tool to investigate both the properties of dark matter haloes and constrain the cosmological parameters.}
 %
{We study how the number of detectable peaks and its scaling with redshift depend upon the cluster dark matter halo profiles and use peak statistics to constrain the parameters of the mass\,-\,concentration (MC) relation. We investigate which constraints the Euclid mission \citep{EditorialTeam:2011mu,Amendola:2012ys} can set on the MC coefficients also taking into account degeneracies with the cosmological parameters.}
%
{To this end, we first estimate the number of peaks and its redshift distribution for different MC relations. We find that the steeper the mass dependence and the larger the normalisation, the higher is the number of detectable clusters, with the total number of peaks changing up to $40\%$ depending on the MC relation. We then perform a Fisher matrix forecast of the errors on the MC relation parameters as well as cosmological parameters.}
%
{We find that peak number counts detected by Euclid can determine the normalization $A_v$, the mass $B_v$ and redshift $C_v$ slopes and intrinsic scatter $\sigma_v$ of the MC relation to an unprecedented accuracy being $\sigma(A_v)/A_v = 1\%$, $\sigma(B_v)/B_v = 4\%$, $\sigma(C_v)/C_v = 9\%$, $\sigma(\sigma_v)/\sigma_v = 1\%$ if all cosmological parameters are assumed to be known. Should we relax this severe assumption, constraints are degraded, but remarkably good results can be restored setting only some of the parameters or combining peak counts with Planck data. This precision can give insight on competing scenarios of structure formation and evolution and on the role of baryons in cluster assembling. Alternatively, for a fixed MC relation, future peaks counts can perform as well as current BAO and SNeIa when combined with Planck.}
 {}


\maketitle

\section{Introduction}

A clear picture of the formation and evolution of cosmic structures requires a good understanding of the interplay between astrophysical processes and the cosmological framework. As dark matter dominated, nearly virialised objects, clusters of galaxies should be relatively easy to sort out. The hierarchical cold dark matter scenario with a cosmological constant ($\Lambda$CDM) can explain many features of galaxy clusters. Their density profile over most radii is accurately reproduced by the Navarro\,-\,Frenk\,-\,White (NFW) density profile \citep{Navarro:1995iw,Navarro:1996gj} and the relation between mass and concentration is accurately predicted.

The concentration measures the halo central density relative to outer regions and is related to the cluster properties at the formation time, in particular to its virial mass and redshift \citep{Bullock:1999he}. Smaller mass and smaller redshift clusters should show higher concentrations, with a moderate evolution with mass and redshift \citep{Bullock:1999he,Duffy:2008pz}. A flattening in concentration might appear towards the very high masses tail and high redshifts \citep{2011ApJ...740..102K,2012MNRAS.423.3018P}, but the real presence of any turn\,-\,around is still debated \citep{Meneghetti:2013nma}.

The mass\,-\,concentration (hereafter, MC) relation also depends upon cosmological parameters \citep{Kwan:2012nd,2013MNRAS.428.2921D}. Other than the normalisation of the matter power spectrum and the dark matter content, the dark energy equation of state also impacts the MC relation for low mass haloes \citep{Kwan:2012nd,2013MNRAS.428.2921D}, although the effect is nevertheless secondary in very massive clusters.

This clear theoretical picture is challenged by conflicting observational pieces of evidence \citep{Comerford:2007xb,Fedeli:2011gj}. The normalisation factor of the MC relation is larger than expected, whereas the slope is steeper \citep{Comerford:2007xb,Ettori:2010di}). These results seem to be stable against redshift. A steep $c(M)$ was found at $0.15\ls z\ls 0.3$ in the weak lensing analysis of $19$ X\,-\,ray luminous lensing galaxy clusters \citep{Okabe:2009pf}, at $0.3\ls z \ls 0.7$ in a combined strong and weak lensing  analysis of a sample of $25$ lenses from the Sloan Giant Arcs Survey \citep{Oguri:2011dt} and in a sample of $31$ massive galaxy clusters at high redshift $0.8\ls z \ls 1.5$ \citep{2013MNRAS.434..878S}. Concentrations measured in lensing selected clusters are systematically larger than in X\,-\,ray analyses \citep{Comerford:2007xb} and a significant number of over\,-\,concentrated clusters is detected at high masses \citep{Broadhurst:2008re,Oguri:2008ww,Umetsu:2011ip}. However, 
these conflicts 
can be partially reconciled by considering orientation and shape biases \citep{2011MNRAS.416.3187S,Rasia:2012jz}, which most severely affect strong lenses. In fact, the disagreement is reduced in strong lensing analyses of X\,-\,ray selected samples \citep{2012MNRAS.419.3280S}. It is also worth noting that the discrepancy can also be related to selection effects. Indeed, the recent analyses of the CLASH data (Merten et al. 2014) have shown that the MC relation determined from the data can be reconciled with the expectation from numerical simulations provided one carefully takes into account the details of the sample selection.

One of the main source of concern in the measurement of the MC relation is the choice and size of the sample. Clusters selected according to their gravitational lensing features or X\,-\,ray flux may form biased samples preferentially elongated along the line of sight \citep{Hennawi:2005bm,2011A&A...530A..17M} and the strongest lenses are expected to be a highly biased population of haloes preferentially oriented towards the observer \citep{Oguri:2008ww}. Neglecting halo triaxiality can then lead to systematically larger biased concentrations. Correcting for shape and orientation biases requires very deep multi wavelength observations \citep{2013MNRAS.428.2241S,Limousin:2012qd}, which are expensive to carry out on a large sample. Furthermore, the orientation bias cannot fully account for the discrepancy between theory and observations for some very strong lenses \citep{2010MNRAS.403.2077S}.

Another reason of concern is that the discrepancies between theory and observations are mitigated when stacking techniques are employed. Weak lensing analyses of stacked clusters of agree with theoretical predictions \citep{Johnston:2007uc,Mandelbaum:2008iz,2013MNRAS.434..878S}. \citet{Oguri:2011dt} found that the concentration measured with a stacked analysis was smaller than what expected from the individual clusters in their sample. \citet{Okabe:2013efa} performed a weak lensing stacked analysis of a complete and volume-limited sample of X\,-\,ray selected galaxy clusters and found shallow density profiles consistent with numerical simulations. The stacked profile of 31 clusters at high redshift was in accordance with theoretical predictions too \citep{2013MNRAS.434..878S}.

Pending the debate on which MC relation has to be trusted upon---numerically motivated or observationally based---and given the systematics connected to the measurements of concentration, it is worth tackling the problem from a different perspective relying on an alternative probe. Weak lensing peaks in the convergence map have recently attracted attention as a powerful tool to find clusters up to very high redshift. While small statistics and difficulties in galaxy shape measurement from ground have limited the application of this technique to present data (but see, e.g., \citealt{Shan:2011kg} and Refs. therein for recent results), the future availability of large galaxy surveys both from ground (e.g., LSST\footnote{{\tt www.lsst.org}}) and space (e.g., Euclid\footnote{{\tt www.euclid-ec.org}}) motivate a renewed interest in this technique---not only as a tool to find clusters, but also as a way to probe the cosmological parameters \citep{Marian:2008fd,2010MNRAS.402.1049D,Kratochvil:2009wh,Marian:2010mh}, 
the theory of gravity \citep{Cardone:2012zn},  primordial non\,-\,Gaussianity (Maturi et al. 2011) and the dark matter halo properties \citep{Bartelmann:2002dh}. Investigating if such a methodology can allow us to discriminate among different MC relations is the aim of the present work.

As already hinted at above, peak number counts also depend on cosmological parameters so that one should take care of degeneracies among the two set of quantities. Actually, the background cosmology can be hold fixed when considering the peak number count dependence on the MC relation, since most of the cosmological parameters play a minor role in determining peak statistics. Alternatively, one can combine peak counts with other probes, as we investigate here when taking the recent Planck covariance matrix as a prior.

The plan of the paper is as follows. In Sect.\,2, we review how the number of detectable weak lensing peaks can be estimated and we also discuss issues concerning the filter choice and the role of the MC relation. The formulae here obtained are then used in Sects\,3 and 4 to infer the number of detectable peaks and its redshift distribution for five different MC relations, thus highlighting how the results strongly depend on the MC parameters. This encouraging outcome suggests that interesting constraints on the mass and redshift depencence and the normalization of the MC relation can be put by future surveys. We therefore devote Sect.\,5 to discuss Fisher matrix forecast for the case of the Euclid survey, also taking degeneracies with cosmological parameters into account and the impact of baryons. Lastly, Sect.\,6 is devoted to conclusions.

\section{Weak lensing peaks}

Being the largest and most massive mass concentrations, galaxy clusters are ideal candidates to lens background galaxies. The spectacular arcs forming when cluster and source are aligned along the line of sight are indeed remarkable an evidence. In less favourable circumstances, clusters anyway generate a shear field which can be reconstructed from the statistical properties of the shape distributions of background galaxies. In shear maps, clusters can then be detected as peaks clearly emerging out of the noise---thus offering an efficient technique for their identification.

As a peak finder, we consider here the aperture mass defined by \citep{Schneider:1996ug}
\begin{equation}
M_{\rm ap}(\theta) = \int{\kappa(\theta) U(\vartheta - \theta) d^2\theta}
= \int{\gamma_t(\theta) Q(\vartheta - \theta) d^2\theta}
\label{eq: mapdef}
\end{equation}
where $\kappa(\theta)$ and $\gamma_t(\theta) = -{\cal{R}}[\gamma(\theta) \exp{(-2 {\rm i} \phi)}]$ are the convergence and the tangential shear at position $\theta = (\vartheta \cos{\phi}, \vartheta \sin{\phi})$, and $U(\vartheta)$, $Q(\vartheta)$ are compensated filter functions related to each other by the integral equation

\begin{equation}
Q(\vartheta) = - U(\vartheta) + \frac{2}{\vartheta^2} \int_{0}^{\vartheta}{U(\vartheta^{\prime}) \vartheta^{\prime} d\vartheta^{\prime}} \ .
\end{equation}
In order to detect a cluster as a peak in the aperture mass map, we have preliminarily to estimate the $M_{\rm ap}$ variance and then set a cut on the signal\,-\,to\,-\,noise ratio, $S/N$. To this end, we have to specify how we compute both the signal, as will be outlined in the two following sections.

\subsection{Halo model}
For given lens and source redshifts $(z_l, z_s)$, the aperture mass depends on the lens mass density profile. Motivated by both simulations of structure formation and observations, we assume that cluster haloes are described by the NFW \citep{Navarro:1995iw,Navarro:1996gj} model
\begin{equation}
\rho(r) = \frac{\rho_s}{x (1 + x)^2} = \frac{(M_{\rm vir}/4 \pi R_{\rm vir}^3) f_{\rm NFW}(c_{\rm vir})}{(c_{\rm vir} y) (1 + c_{\rm vir} y)^2}
\label{eq: rhonfw}
\end{equation}
with $x = r/R_s$, $y = r/R_{\rm vir}$ and
\begin{equation}
f_{\rm NFW}(c_{\rm vir}) = \frac{c_{\rm vir}^3}{\ln{(1 + c_{\rm vir})} - c_{\rm vir}/(1 + c_{\rm vir})} \ .
\label{eq: fc}
\end{equation}
Following a common practice, we use as a mass parameter the virial mass $M_{\rm vir}$, i.e. the mass lying within the virial radius $R_{\rm vir}$, where the mean mass density, $\bar{\rho} = M_{\rm vir}/(4/3) \pi R_{\rm vir}^3$, equals $\Delta_{\rm vir}(z_l) \rho_{\rm crit}(z_l)$ (with $\rho_{\rm crit}(z_l)$ the critical density at the cluster redshift). The critical overdensity $\Delta_{\rm vir}(z_l)$ should be computed according to the spherical collapse formalism (being hence a function of the adopted cosmological model) and turns out to depend on the redshift (see, e.g. \citealt{Bryan:1997dn}). However, it is not unusual to set $\Delta_{\rm vir} = 200$ at all redshift and replace $(M_{\rm vir}, c_{\rm vir})$ with $(M_{200}, c_{200})$ which is what we will do in the following.

As a second parameter, we choose the halo concentration\footnote{Hereafter, with an abuse of terminology, we will refer to $(M_{200}, c_{200})$ as the virial mass and concentration, although formally this definition applies to $(M_{\rm vir}, c_{\rm vir})$ only.} $c_{200} = R_{200}/R_s$. According to $N$body simulations, the NFW model can be reduced to a one parameter class since $c_{200}$ correlates with the virial mass $M_{200}$. Actually, the slope, the scatter and the redshift evolution of the $c_{200}$\,-\,$M_{200}$ relation are still matter of controversy, with different results available in the literature. However, most of works do agree on the shape of the MC relation given by
\begin{equation}
c_{200}(M_{200}, z_l) = A_{v} \left ( \frac{M_{200}}{M_{\rm piv}} \right )^{B_v} (1 + z_l)^{C_v}
\label{eq: mclaw}
\end{equation}
with $M_{\rm piv}$ a pivot mass and different values for the $(A_v, B_v, C_v)$ parameters. Eq.~(\ref{eq: mclaw}) should actually be considered only as an approximate rather than exact relation. Indeed, for a given mass, halo concentrations scatter around the value predicted by Eq.~(\ref{eq: mclaw}). This gives rise to a distribution which can be reasonably well described as a lognormal with mean value and variance $\sigma_v$ depending on the MC relation adopted. Such a scatter is usually neglected in peak count studies, but it must be taken into account if one aims at studying the impact of the MC relation on peaks statistics. We set $M_{\rm piv} = 5 \times 10^{14} \ {\rm M_{\odot}}$ and
\begin{equation}
(A_v, B_v, C_v, \sigma_v) = (3.59, -0.084, -0.47, 0.15)
\end{equation}
as a fiducial case in agreement with \citet[][hereafter D08]{Duffy:2008pz}. 

With these ingredients, it is now only a matter of algebra to compute the lensing properties of the NFW profile inserting the mass and concentration values in the analytical expressions of the convergence $\kappa$ and shear $\gamma$ given in \cite{Bartelmann:1996hq} and \cite{2000ApJ...534...34W}.

\subsection{Filter function and $S/N$ ratio}
Gravitational lensing probes the total matter distribution along the line of sight so that the observed aperture mass $M_{\rm ap}$ is eventually the sum of the cluster contribution as well as another due to the uncorrelated large-scale structure projected along the same line of sight, namely $M_{\rm ap} = M_{\rm ap}^{\rm clust} + M_{\rm ap}^{\rm LSS}$. Being a density contrast, one typically assumes that $M_{\rm ap}^{\rm LSS}$ averages out to zero---thus not biasing the $M_{\rm ap}$ distribution, but only contributing to the the variance \citep{2001A&A...370..743H}. Then, the filter functional form and its parameters are chosen as a compromise between the need to find as many clusters as possible and the necessity of decreasing the number of fake peaks. To this end, shear field simulations, taking into account both the underlying fiducial cosmology and the survey characteristics, are used to tailor the filter parameters (see e.g. \citealt{2005A&A...442...43H}). However, such a method is far from being 
perfect, as it is intimately related to the adopted cosmology and the halo model assumed in the reference simulation.

As a possible solution, \citet[][hereafter M05]{2005A&A...442..851M} proposed an {\it optimal filter} taking explicitly into account both the cosmological model and the shape of the signal, i.e. the halo shear profile. According to M05, the Fourier transform of the filter reads
\begin{equation}
\hat{\Psi}(\ell) = \frac{1}{(2 \pi)^2} \left [ \int{\frac{|\hat{\gamma}_t(\ell)|^2}{P_N(\ell)} d^2 \ell} \right ]^{-1}
\frac{\hat{\gamma}_t(\ell)}{P_N(\ell)} \ ,
\label{eq: defpsihat}
\end{equation}
where $\hat{\gamma}_t$ is the Fourier transform of the tangential shear component and $P_N(\ell)$ the noise power spectrum as a function of the angular wavenumber $\ell$. Two terms contribute to the noise so that it is\,
\begin{equation}
P_N(\ell) = P_{\varepsilon} + P_{\gamma}(\ell) \ ,
\label{eq: defpntot}
\end{equation}
with
\begin{equation}
P_{\varepsilon} = \frac{1}{2} \frac{\sigma_{\varepsilon}^2}{n_g}
\label{eq: defpeps}
\end{equation}
the term due to the finite number of galaxies (with number density $n_g$) and their intrinsic ellipticities with variance $\sigma_{\varepsilon}$; and $P_{\gamma}(\ell) = P_{\kappa}(\ell)/2$ the noise due to the LSS, where the factor $1/2$ originates from using only one shear component. Under the Limber flat sky approximation, we have
\begin{equation}
P_{\kappa}(\ell) =
\left ( \frac{3 \Omega_M H_0^2}{2 c^2} \right )^2 \int_{0}^{\chi_h}{P_{\delta}\left ( \frac{\ell}{\chi}, \chi \right ) \frac{{\cal{W}}^2(\chi)}{a^2(\chi)} d\chi}
\label{eq: limber}
\end{equation}
with
\begin{equation}
P_{\delta}(k, z) = {\cal{A}}_s k^{n_s} {\cal{T}}_M^2(k) D^2(z)
\label{eq: defpdelta}
\end{equation}
the matter power spectrum, with spectral index $n_s$ and present day mass variance $\sigma_8$ on scales $R = 8 h^{-1} \ {\rm Mpc}$ used to set the normalization constant ${\cal{A}}_s$. Here, ${\cal{T}}_M(k)$ is the matter transfer function, approximated according to \citet{Eisenstein:1997ik}, while $D(z)$ is the growth factor (normalized to unity today) for the assumed cosmological model. In Eq.~(\ref{eq: limber}), the redshift is replaced by the comoving distance
\begin{equation}
\chi(z) = c\int_{0}^{z}{\frac{dz^{\prime}}{H(z^{\prime})}}
\label{eq: defchi}
\end{equation}
with $H(z)$ the Hubble rate. 
 Finally, ${\cal{W}}(\chi)$ is the lensing weight function (assuming a spatially flat universe)

\begin{equation}
{\cal{W}}(\chi) = \int_{\chi}^{\chi_h}{\left ( 1 - \frac{\chi}{\chi^{\prime}} \right ) p_{\chi}(\chi^{\prime}) \chi^{\prime} d\chi^{\prime}}
\label{eq: lensweight}
\end{equation}
and $p_{\chi}(\chi) d\chi = p_{z}(z) dz$ the source redshift distribution---which we here parameterise as \citep{1994MNRAS.270..245S}
\begin{equation}
p_z(z) \propto\frac{\beta}{z_0} \left ( \frac{z}{z_0} \right )^2 \exp{\left [ - \left ( \frac{z}{z_0} \right )^{\beta} \right ]}
\label{eq: pz}
\end{equation}
and normalise to unity. We set $(\beta, z_0) = (1.5, 0.6)$ so that $z_m = 0.9$ is the median redshift of the sources as expected for the Euclid mission \citep{Marian:2010mh}.

Following \citet{Bartelmann:1996hq} and \citet{2000ApJ...534...34W} for the shear profile of the NFW model, one finally gets for the Fourier transform of the filter\footnote{Note that the minus sign comes out from our convention on the sign of the tangential shear component.}\,
\begin{multline}
\hat{\Psi}(\ell) =- \frac{1}{(2 \pi)^3} \left [ \left (  \frac{M_{200}/4 \pi R_{200}^2}{\Sigma_{\rm crit}} \right ) g(c_{200})
\left ( 2 \pi \theta_s^2 \right ) \right ]^{-1} \\
 \times \frac{\tilde{\tau}(\ell \theta_s)}{P_N(\ell) \tilde{{\cal{D}}}(\theta_s)}
\label{eq: fourierpsi}
\end{multline}
where $\Sigma_{\rm crit} = c^2 D_s/(4 \pi G D_d D_{ds})$ is the critical density for lensing (depending on the lens and source redshift), $g(c_{\rm vir}) = f_{\rm NFW}(c_{200})/c_{200}$,  $\theta_s$ is the angular scale corresponding to $R_s$, and we have defined
\begin{align}
\tilde{\tau}(\ell \theta_s) &= \int_{0}^{\infty}{\tilde{\gamma}(\xi) J_2(\ell \theta_s \xi) \xi d\xi} \ ,
\label{eq: deftautilde}\\
\tilde{{\cal{D}}}(\theta_s) &= \int_{0}^{\infty}{\frac{\left | \tilde{\tau}(\ell \theta_s) \right |^2}{P_N(\ell)} \ell d\ell} \ ,
\label{eq: defdtilde}
\end{align}
with $\tilde{\gamma}(\xi)$ the NFW shear profile scaled with respect to $\rho_s R_s /\Sigma_{\rm crit}$ \citep{Bartelmann:1996hq,2000ApJ...534...34W}.

Let us now stress an important caveat about the values of $(M_{200}, c_{200})$ entering the filter function. The best choice would be to fix them to those of the halo we are interested in to find. Needless to say, such a strategy is unfeasible and we have to set them to some fiducial values $(M_{200}^{\rm fid}, c_{200}^{\rm fid})$, thus obtaining a filter optimised for finding clusters with mass and concentration close to the fiducial ones. As we will see in a moment, the $S/N$ ratio does not critically depend on these values, but rather on the MC relation taken as fiducial.

With this caveat in mind, we take the inverse Fourier transform of Eq.~(\ref{eq: fourierpsi}) and set $Q = \Psi$ in the aperture mass definition to get eventually
\begin{equation}
M_{\rm ap}(\vartheta; \theta_s) = \frac{1}{(2 \pi)^4} \frac{M_{200}}{M_{200}^{\rm fid}} \left ( \frac{R_{200}^{\rm fid}}{R_{200}} \right )^2
\frac{g\left (c_{200} \right )}{g\left (c_{200}^{\rm fid} \right )} \frac{\tilde{M}_{\rm ap}\left (\vartheta, \theta_s/\theta_s^{\rm fid} \right )}{\tilde{{\cal{D}}}\left (\theta_s^{\rm fid} \right )},
\label{eq: endmap}
\end{equation}
with $\vartheta$ the filter aperture and the label `fid' denoting quantities referred to the fiducial case. In Eq.~(\ref{eq: endmap}), we have also defined (with $\xi = \vartheta/\theta_s^{\rm fid}$ and $\xi^{\prime} = \theta/\theta_s^{\rm fid}$)
\begin{multline}
\tilde{M}_{\rm ap} = \int_{0}^{\xi}{\tilde{\gamma}\left [ \xi^{\prime} \left ( \theta_s/\theta_s^{\rm fid} \right ) \right ] \xi^{\prime} d\xi^{\prime}} \\\times
\int_{0}^{2 \pi}{\tilde{\Psi}\left [ \theta_s \left ( \xi^2 + \xi^{\prime 2} - 2 \xi \xi^{\prime} \cos{\theta} \right )^{1/2} \right ]
\cos{(2 \theta)} d\theta}
\label{eq: defmaptilde}
\end{multline}
and $\tilde{\Psi} = \tilde{\tau}\left (\ell \theta_s^{\rm fid} \right )/P_N(\ell)$.

A conceptual remark is in order here. Eq.~(\ref{eq: endmap}) has been obtained assuming that the measured shear and cluster profiles are the same. Actually, what one measures is the shear field reconstructed from the observed galaxies ellipticities, which is only an approximation of the input cluster profile. However, the only way to get analytic prediction is to identify the reconstructed and theoretical shear profiles---what we have implicitly done here. Moreover, as a further approximation, we have set $\gamma \simeq g_{\rm sh}$ with $g_{\rm sh} = \gamma/(1 - \kappa)$ the measurable reduced shear. In the weak lensing limit ($\kappa << 1$), the difference is safely negligible.

In order to predict the number of peaks, we need the $S/N$ ratio. Then, we first compute the noise given by\footnote{Note that, contrarily to the usual procedure, we are including only $P_{\varepsilon}$ as noise term rather than the full one $P_{N}(\ell)$. This is motivated by our definition of the signal as the sum of the cluster and LSS peaks so that only $P_{\varepsilon}$ has to be considered as noise.} \citep{2005A&A...442..851M}
\begin{align}
\sigma_{\rm ap}^2 & = \frac{1}{2 \pi} \int_{0}^{\infty}{P_{\varepsilon} \left | \tilde{\Psi}(\ell) \right |^2 \ell d\ell} =\nonumber \\
 & = \frac{1}{(2 \pi)^7} \left ( \frac{M_{200}^{\rm fid}/4 \pi R_{200}^{fid \ 2}}{\Sigma_{\rm crit}} \right )^{-2} \frac{g^{-2}\left (c_{200}^{\rm fid} \right )}
{\left ( 2 \pi \theta_s^{fid \ 2} \right )^2} \frac{P_{\varepsilon}}{\tilde{{\cal{D}}}(\theta_s^{\rm fid})} \nonumber \\
 &\times \int_{0}^{\infty}{\frac{\tilde{\tau}^2(\ell \theta_s^{\rm fid})}{[P_{\varepsilon} + (1/2) P_{\kappa}(\ell)]^2} \ell d\ell}.
\label{eq: defsigmaap}
\end{align}
Thus, the $S/N$ ratio reads
\begin{equation}
{\cal{S}}(\vartheta; {\bf p}) = \frac{1}{\sqrt{2 \pi}}  \frac{M_{200}/4 \pi R_{200}^2}{\Sigma_{\rm crit}}
\frac{2 \pi \theta_s^2 g(c_{200})}{\sqrt{P_{\varepsilon}}} \frac{\tilde{M}_{\rm ap}(\vartheta)}{\tilde{\sigma}_{\rm ap}},
\label{eq: stonvsz}
\end{equation}
where $\tilde{\sigma}_{\rm ap}$ is the integral in the third row of Eq.~(\ref{eq: defsigmaap}). In the above relation, ${\bf p}$ summarizes the parameters which the $S/N$ ratio depends upon. Apart from the virial mass $M_{200}$ explicitly appearing as a multiplicative term through the factor $M_{200}/R_{200}^2 \propto M_{200}^{1/3}$, there are the MC relation parameters $(A_v, B_v, C_v)$ used to estimate the halo concentration $c_{200}$ and hence $\theta_s$ through Eq.~(\ref{eq: mclaw}). The lens and source redshift $(z_l, z_s)$ enter through the critical density $\Sigma_{\rm crit}$ and the conversion from the linear $R_s$ to the angular scale $\theta_s$. Finally, we remember that Eq.~(\ref{eq: mclaw}) is affected by a lognormal scatter $\sigma_v$ which is another parameter to be added to the list. In order to take into full account both the source redshift distribution and the scatter in the MC relation, we therefore compute the final $S/N$ ratio as
\begin{equation}
{\cal{S}}(\vartheta; z_l, M_{200}) =
\int_{z_l}^{\infty}{\tilde{{\cal{S}}}(\vartheta; z_l, z_s, M_{200}) p_z(z_s) dz_s},
\label{eq: stonend}
\end{equation}
with
\begin{equation}
\tilde{{\cal{S}}}(\vartheta; z_l, z_s, M_{200}) =
\int{{\cal{S}}(\vartheta; {\bf p}) p_c(c_{200}, M_{200}) dc_{200}}
\label{eq: stonpreend}
\end{equation}
and
\begin{equation}
p_c(c_{200}; M_{200}) \propto \exp{\left \{- \frac{1}{2} \left [ \frac{\log{c_{200}} - \log{\langle c_{200} \rangle(M_{200})}}{\sigma_v} \right ]^2 \right \}}
\label{eq: pclog}.
\end{equation}
Here, $\langle c_{200} \rangle(M_{200})$ as in Eq.~(\ref{eq: mclaw}) for given values of the MC relation parameters $(A_v, B_v, C_v)$ and scatter $\sigma_v$.

Note that two sets of MC relation parameters actually enter Eqs~(\ref{eq: stonend})\,-\,(\ref{eq: pclog}), the former fixed by the MC relation used to set the filter (hence determining $\theta_s^{\rm fid}$) and the latter related to the MC relation used in the estimate of the expected number of clusters (giving $\theta_s)$. We will use the D08 MC relation as a fiducial (with $\sigma_v = 0.15)$ to set the filter function, while we consider different choices for $(A_v, B_v, C_v, \sigma_v)$ to investigate whether peak statistics can discriminate among MC relations.

\section{Cluster detectability}
Eqs~(\ref{eq: stonvsz})\,-\,(\ref{eq: pclog}) allow us to estimate the $S/N$ ratio for a cluster of virial mass $M_{200}$ and redshift $z_l$ provided the MC relation parameters $(A_v, B_v, C_v)$ and the scatter $\sigma_v$. To this end, we have to set preliminarily the survey characteristics and the background cosmology to estimate the noise $\sigma_{\rm ap}$. Moreover, we have also to fix the filter scale $\vartheta$.

We consider the survey specification for the photometric Euclid survey \citep{EditorialTeam:2011mu,Amendola:2012ys}, with an area of $15000 \ {\rm deg^2}$ and an ellipticity dispersion $\sigma_{\epsilon}= 0.3$. The total number of source galaxies is set to $n_g = 30 \ {\rm gal/arcmin}^2$ and assumed to be uniform over the full survey area. The results can be easily scaled to other choices noting that ${\cal{S}} \propto \sqrt{n_g}$, while the total number of peaks linearly depends on the survey area.

In order to be consistent with the recent Planck results \citep{2013arXiv1303.5076P}, we assume a flat $\Lambda$CDM model as fiducial cosmological scenario with
\begin{equation}
(\Omega_M, \Omega_b, h, n_s, \sigma_8) = (0.306, 0.048, 0.678, 0.961, 0.826) \ ,
\end{equation}
where $\Omega_M$ ($\Omega_b$) is the present day matter (baryon) density, $h = H_0/100 \ {\rm km/s/Mpc}$ the present-day dimensionless Hubble constant, $n_s$ the scalar spectral index and $\sigma_8$ the variance of perturbations on the scale $8 h^{-1} \ {\rm Mpc}$. The dark energy equation of state is described by the CPL \citep{Chevallier:2000qy,Linder:2002et} ansatz
\begin{equation}
w = w_0 + w_a (1 - a) = w_0 + w_a z/(1 + z)
\end{equation}
with $(w_0, w_a)$ held fixed to the $\Lambda$CDM values, viz. $(-1, 0)$. We also assume dark energy does not cluster on the scales of interest. Thus, the growth factor is solution of
\begin{equation}
\ddot{\delta} + 2 H \dot{\delta} - 4 \pi G \rho_M \delta = 0
\end{equation}
with $\delta$ the density perturbation. The $S/N$ ratio has a negligible dependence on the cosmological parameters, so that the following results hold true independently on the fiducial cosmology adopted.

The choice of the filter scale $\vartheta$ asks for some caution. The optimal filter is designed accounting for the NFW profile to maximise the signal. Therefore, a natural scale would be $\vartheta = \theta_s$, as most of the mass contributing to the lensing signal is contained within this aperture. For a cluster with $M_{200} = 5 \times 10^{14} \ {\rm M_{\odot}}$ at a typical cluster redshift $z_l = 0.3$, the concentration predicted by the D08 relation reads $c_{200}  = 2.65$ thus giving $\theta_s \simeq 1 \ {\rm arcmin}$, while the virial radius subtends an angle $\theta_{200} \simeq 2.8 \ {\rm arcmin}$. However, not all the clusters have the same mass and redshift. Setting $\vartheta = 1 \ {\rm arcmin}$ would be an optimal choice for these median values, but would strongly underestimate the signal for clusters more massive or at a lower redshift. On the contrary, a varying $\vartheta$ would allow to maximise the $S/N$ at every redshift, but would make it difficult to compare peak counts in different bins.
 As a compromise, we therefore set $\vartheta = 2 \ {\rm arcmin}$ noting that with this choice we cut the contribution on scales larger than $\theta_{200}$. Indeed, it can be smaller than $2 \ {\rm arcmin}$ for small mass and/or high redshift clusters.

Having set all the preliminary quantities, we can now investigate how the $S/N$ ratio depends on the virial mass and redshift for different MC relations. Table~\ref{tab: mclist} lists the parameter choices for different models we take as representative cases, also giving the id which we will use in the following.
\begin{table}
\centering
\caption{\label{tab: mclist}MC relations parameters for a pivotal mass $M_{\rm piv} = 5 \times 10^{14} h^{-1} \ {\rm M_\odot}$ for the different cases investigated, whose ids are shown in the first column. The scatter is set to $\sigma_v = 0.15$ for the D08 relation and to $\sigma_v = 0.12$ for other models in agreement with what reported in the literature. Note that the D08 relation is the one used to compute the filter.}
\begin{tabular}{lcccl}
\hline
Id & $A_v$ & $B_v$ & $C_v$ & Reference \\
\hline \hline
~ & ~ & ~  \\
B01 & $5.70$ & $-0.13$ & $-1.00$ & \citet{Bullock:1999he} \\
D08 & $3.59$ & $-0.085$ & $-0.47$ & \citet{Duffy:2008pz}\\
Ok10 & $4.60$ & $-0.40$ & $0.00$ & \citet{Okabe:2009pf} \\
Ok10z & $4.60$ & $-0.40$ & $-0.47$ & This paper \\
Og12 & $7.70$ & $-0.59$ & $0.00$ & \citet{Oguri:2011dt} \\
Og12z & $7.70$ & $-0.59$ & $-0.47$ & This paper \\
\hline
\end{tabular}
\end{table}

The first two cases are theoretical relations motivated by the comparison with numerical simulations. On the contrary, the Ok10 and Og12 MC relations have been inferred from observations\footnote{Actually, \cite{Oguri:2011dt} estimated the $c_{\rm vir}$-$M_{\rm vir}$ rather than the $c_{200}$-$M_{200}$ relation so that the parameters in Table~\ref{tab: mclist} should be changed to take into account this difference. We have, however, neglected this correction.}. The authors did not try to fit for a redshift dependence so that the parameter $C_v$ is set to zero. In order to explore a possible redshift dependence, we have therefore introduced the two further relations (Ok10z and Og12z) by arbitrarily setting the $C_v$ parameter to the D08 value. Although the Ok10z and Og12z relations are not motivated by either numerical simulations or observations, they are useful to scrutinise the dependence of peak statistics on the MC relation. Numerically inspired B01 and D08 relations are shallower than the 
observationally motivated Ok10 and Og12, and have as well a larger normalization. In other words, for the pivot mass, a cluster at a given redshift $z_l$ is more concentrated according to Ok10 and Og12 than for B01 and D08. More concentrated haloes have larger masses within the scale radius, one should thence expect that, for given $(z_l, M_{200})$, the $S/N$ will be larger for Ok10 and Og12 MC. However, this is only partly true. Indeed, due to the different scaling with $z$ of the considered MC relations and the use of a filter based on D08, the $S/N$ will not simply scale with the concentration so that a full computation is needed to check the above qualitative prediction.

\begin{figure*}
\resizebox{\hsize}{!}
{\includegraphics[width=4.5cm]{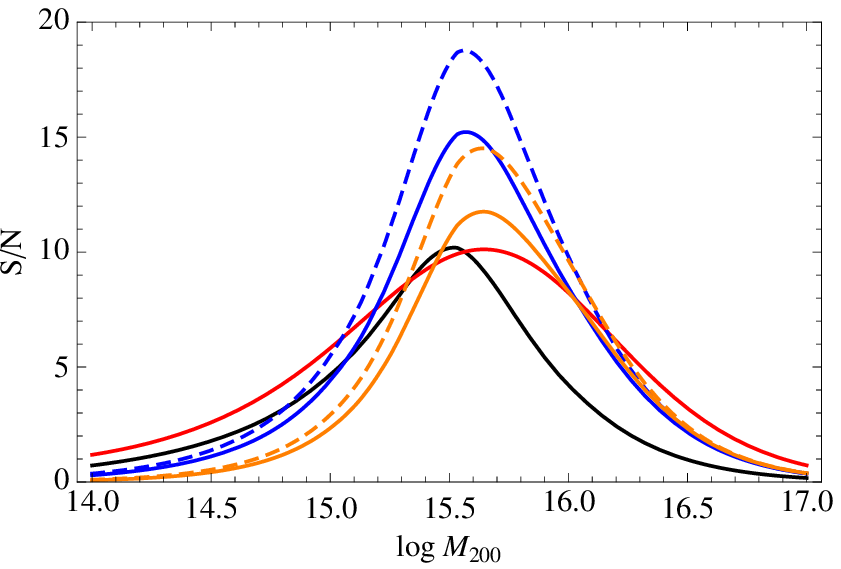}
\includegraphics[width=4.5cm]{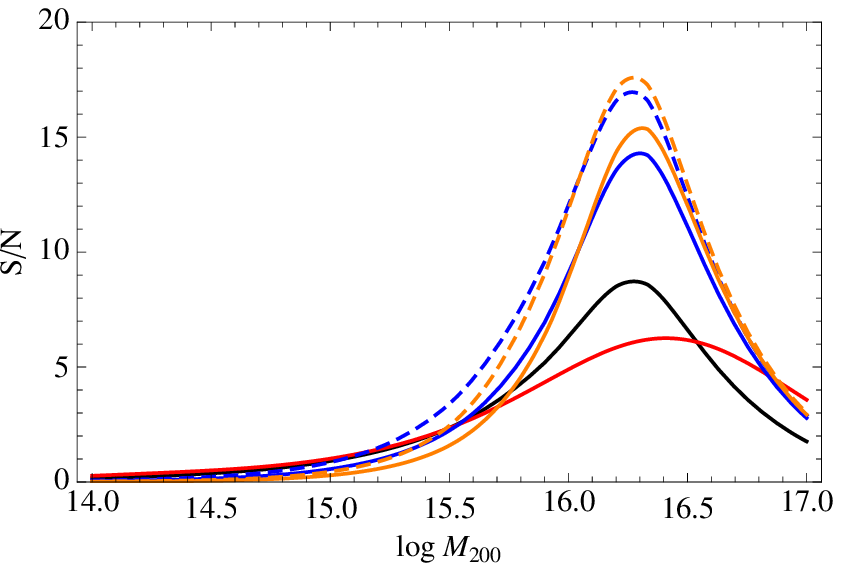}
\includegraphics[width=4.5cm]{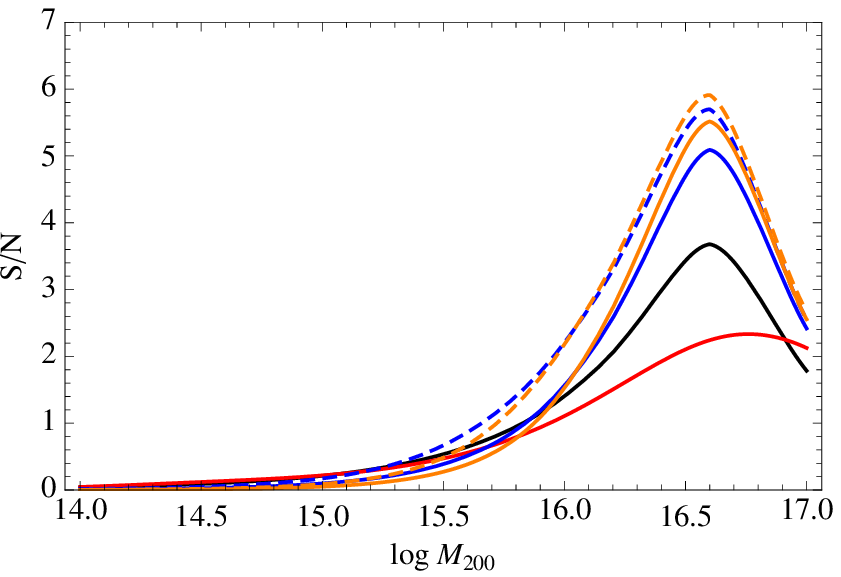}}
\resizebox{\hsize}{!}
{\includegraphics[width=4.5cm]{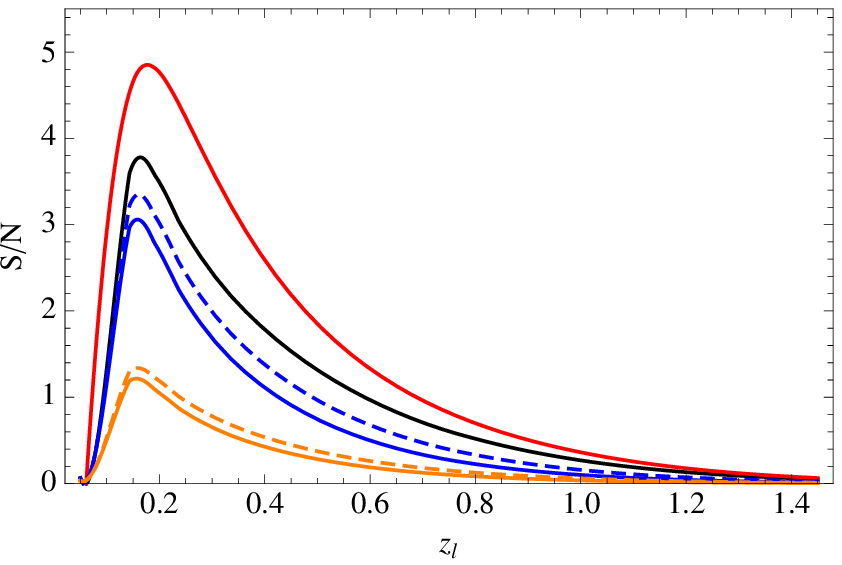}
\includegraphics[width=4.5cm]{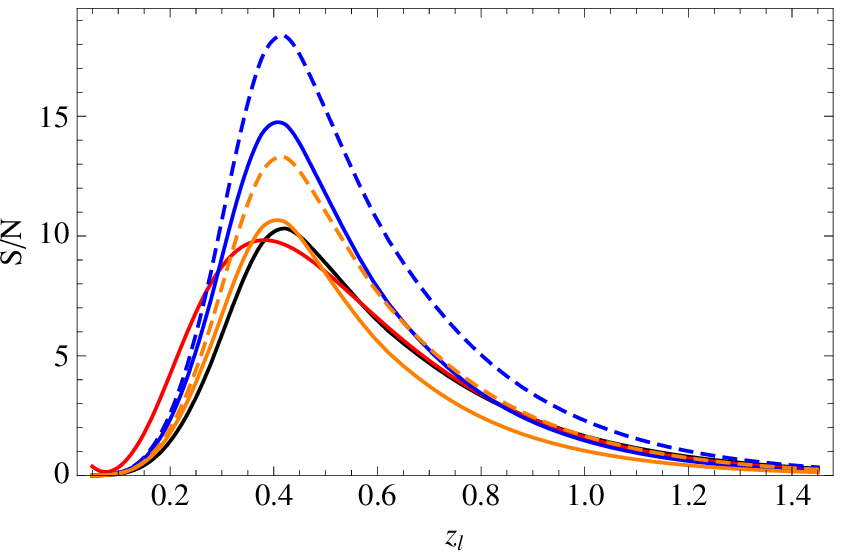}
\includegraphics[width=4.5cm]{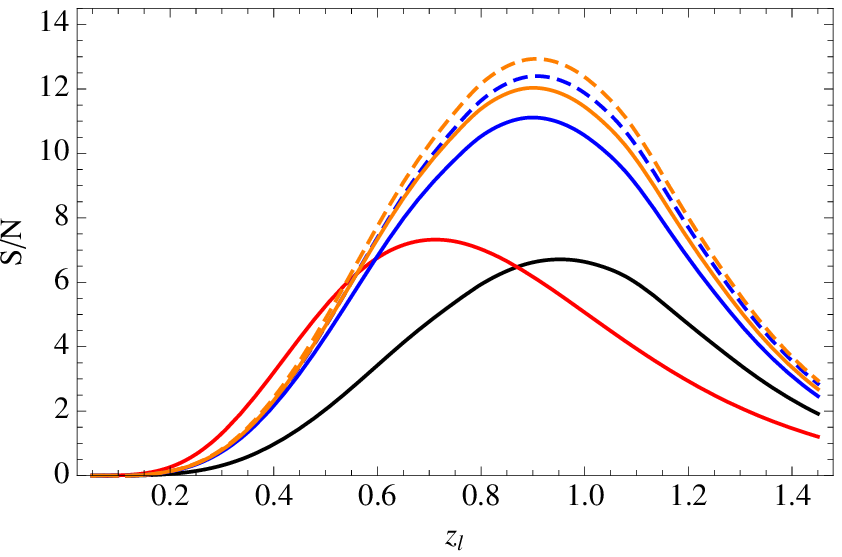}}
\caption{{\it Top panels.} $S/N$ vs cluster virial mass for fixed redshift ($z_l = 0.4, 0.9, 1.3$ from left to right) for different MC relations (black, red, blue, dashed blue, orange, dashed orange for B01, D08, Ok10, Ok10z, Og12, Og12z, respectively). {\it Bottom panels.} $S/N$ vs cluster redshift for fixed mass ($\log{M_{200}} = 14.5, 15.5, 16.5$ from left to right).}
\label{fig: stonplots}
\end{figure*}
\begin{figure*}
\resizebox{\hsize}{!}
{\includegraphics[width=4.5cm]{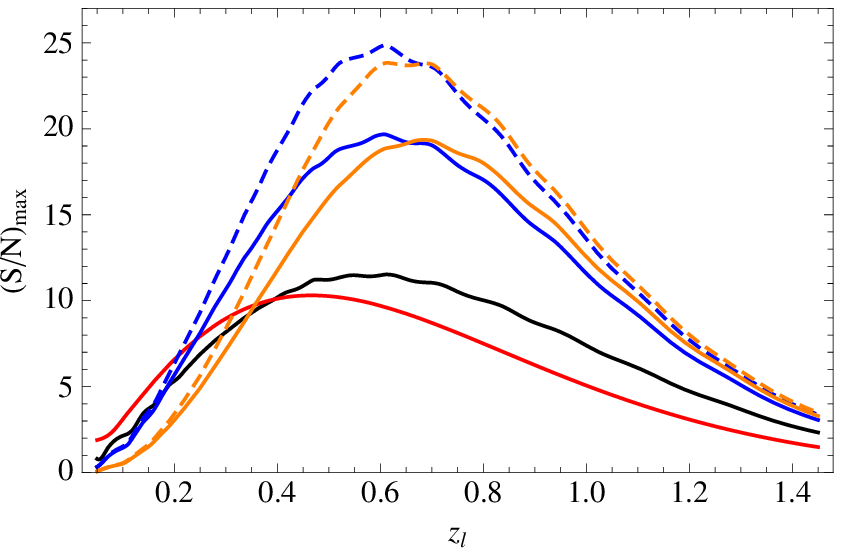}
\includegraphics[width=4.5cm]{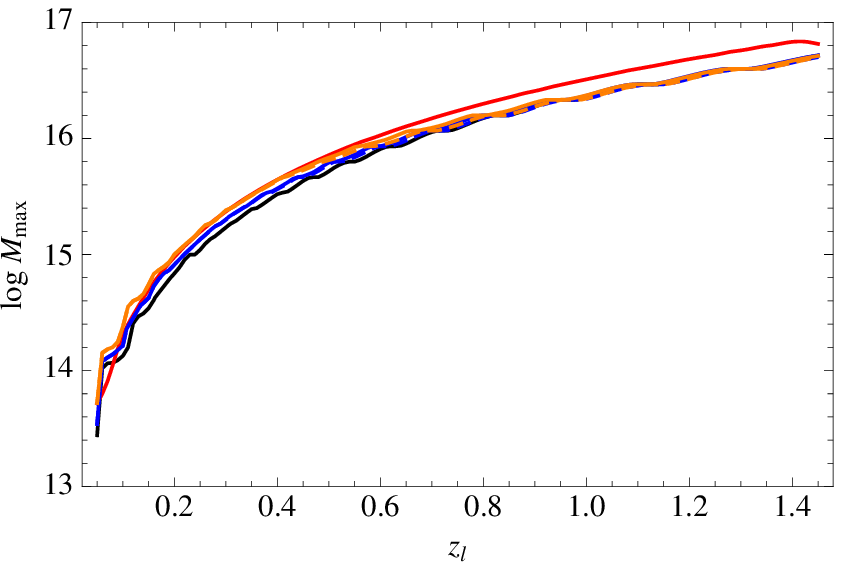}
\includegraphics[width=4.5cm]{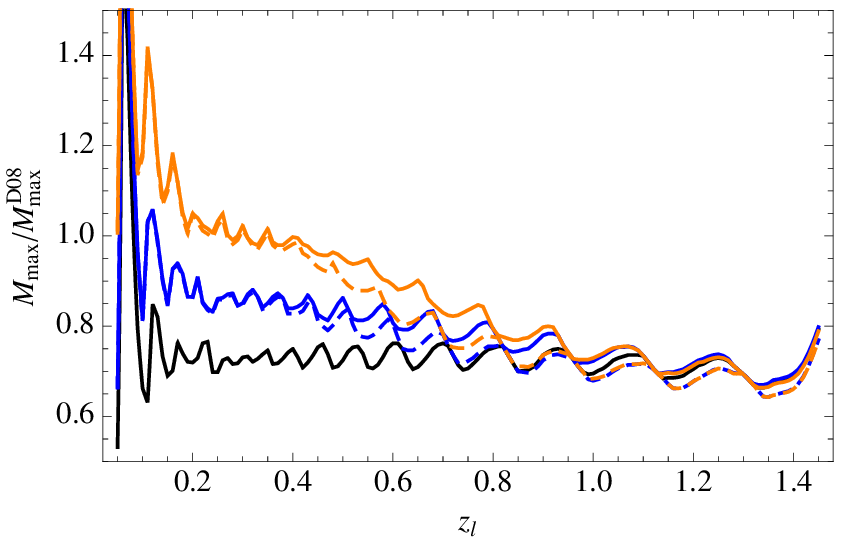}}
\caption{Maximum $S/N$ (left) and virial mass $\log{M_{max}}$ (centre) at which the $S/N$ is maximised as a function of the cluster redshift. The right panel shows $M_{max}$ normalised to the value for the D08 relation (wiggles are merely interpolation features). Black, red, blue, dashed blue, orange and dashed orange lines refer to B01, D08, Ok10, Ok10z, Og12 and Og12z, respectively.}
\label{fig: maxston}
\end{figure*}

Fig.~\ref{fig: stonplots} shows how the $S/N$ ratio ${\cal{S}}$ depends on $(z_l, M_{200})$ for the different MC relations we consider. Although the filter is built using D08 as a fiducial case, the MC relation which gives the largest ${\cal{S}}$ value depends on $(z_l, M_{200})$. As a general rule, for a given $z_l$, the ${\cal{S}}$ vs $M_{200}$ curve has a non-monotonic behaviour. It first increases with $M_{200}$ up to a maximum value and then decreases again. The peak value is larger for steeper MC relations, while the opposite is observed for what concerns the width of the curve. As a consequence, the D08 relation provides the largest ${\cal{S}}$ values for groups and intermediate mass clusters, while the empirically motivated MC relations Ok10 and Og12 (and their redshift dependent counterparts) overcome D08 in the large mass regime.

For fixed cluster mass, the dependence on the redshift is more complicated and which MC relation provides the largest $S/N$ depends on the mass regime. This can also be understood from Fig.~\ref{fig: maxston}, where we plot the maximum $S/N$ ratio as a function of the cluster redshift. As a further remark, we note that the Ok10z and Og12z curves stay always quite close to their redshift independent counterparts, thus demonstrating that it is the concentration mass dependence what drives the ${\cal{S}}$ values.

The above results can be qualitatively explained considering how ${\cal{S}}$ depends on the halo concentration. On the one hand, Eq.~(\ref{eq: stonend}) shows that the $S/N$ ratio is the product of $\theta_s^2 g(c_{200})$, an increasing function of the concentration, and of an integral depending on the ratio $\theta_s/\theta_s^{\rm fid}$ and the filter aperture. One can naively expect that the larger the concentration, the larger the $S/N$ ratio. Should this be the dominant factor, the MC relation providing the larger concentration would also be the one preferred by a $S/N$ viewpoint. However, a larger $c_{200}$ also implies a smaller $\theta_s$. Consequently, if $\theta_s < \vartheta$, the filter cuts away a large part of the cluster, thus leading to a lower aperture mass (and hence a smaller integral term). The best compromise between these two somewhat opposite behaviours depend on the cluster mass and redshift.

It is worth noting that the largest $S/N$ does not guarantee the final detected number of peaks to be also the largest one. This can be understood by looking at Fig.~\ref{fig: maxston}, where we plot the maximum $S/N$ as a function of $z_l$ for the different MC relations considered. For sources at the survey median redshift $z_s = 0.9$, all the MC relations achieve a maximum $S/N$ larger than that of the D08 case. However, such a maximum corresponds to haloes as massive as $\log{M_{200}} \sim 16$ which are few. If we limit our attention to the mass range corresponding to $0.1 \le z_l \le 0.3$, i.e. $13.5 \le \log{M_{200}} \le 15.5$, we see that the D08 maximum $S/N$ is comparable to---if not even larger than---those of other MC relations. Albeit such a discussion only considers the maximum $S/N$, it nevertheless warns against inferring any conclusion on which MC relation provides the largest number of peaks based on $S/N$ only.

\section{Peak number counts}

\begin{figure*}
\resizebox{\hsize}{!}
{\includegraphics[width=4.5cm]{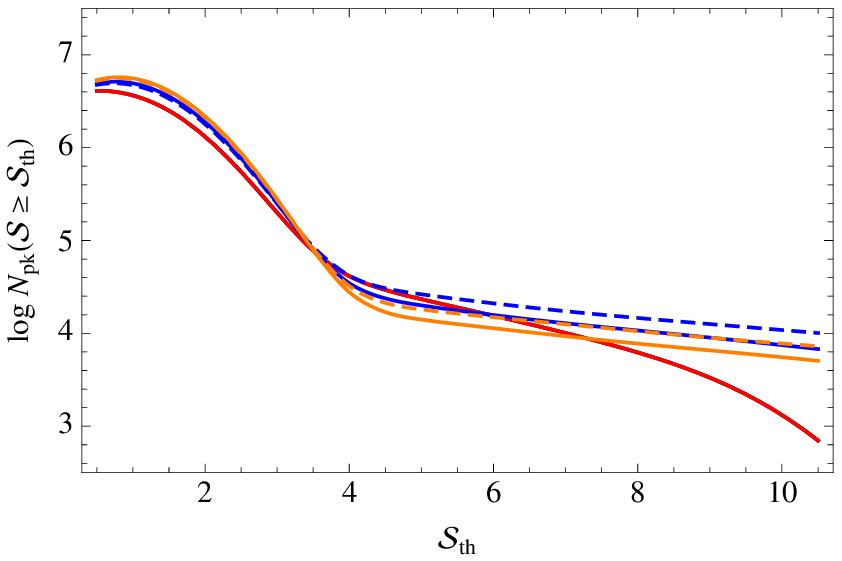}
\includegraphics[width=4.5cm]{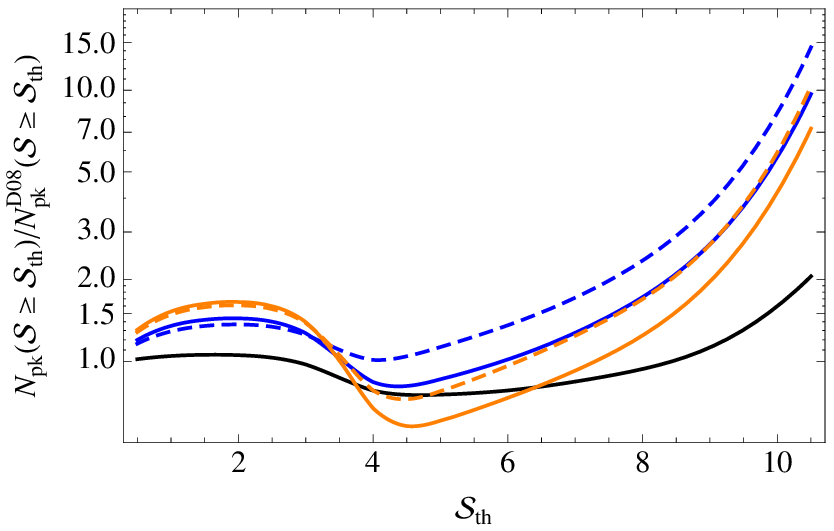}
\includegraphics[width=4.5cm]{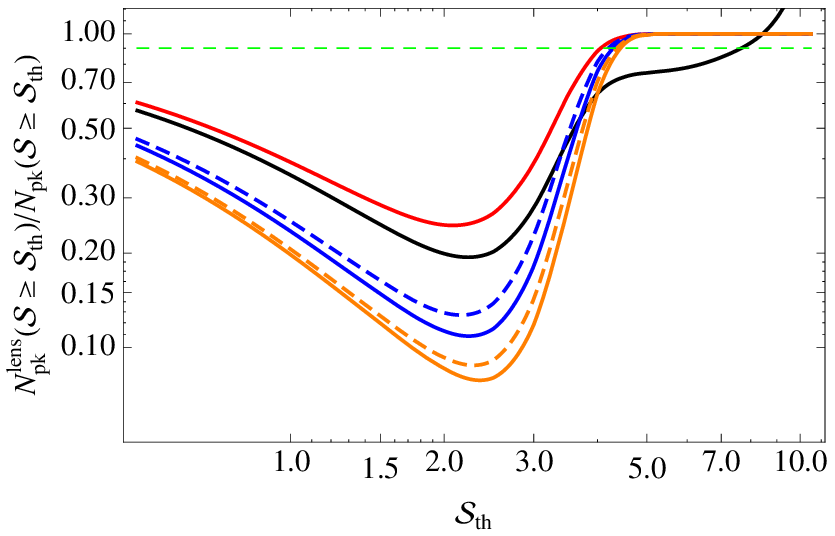}}
\caption{{\it Left.} Total number of peaks as a function of the threshold $S/N$ ratio ${\cal{S}}_{\rm th}$. {\it Centre.} Same as before, but scaled with respect to the D08 fiducial MC relation. {\it Right.} Ratio between the number of peaks due to clusters and the total one as a function of ${\cal{S}}_{\rm th}$. In each panel, black, red, blue, dashed blue, orange and dashed orange lines refer to the B01, D08, Ok10, Ok10z, Og12, Og12z MC relations.}
\label{fig: npktot}
\end{figure*}

The weak lensing peaks we are interested in are due to massive clusters. In order to compute their expected number, we first need to know how many massive haloes there are. This is given by the mass function\footnote{Hereafter, we drop the labels $200$ from the mass and $l$ from the lens redshift to shorten the notation.}
\begin{equation}
{\cal{N}}(\ln{M}) = \frac{\rho_M(z = 0)}{M}  \ \frac{d\ln{\nu}}{d\ln{M}}  \ \nu \varphi(\nu) \  .
\label{eq: mf}
\end{equation}
Here, $\nu = \delta_c/\sigma(M)$, $\delta_c$ is the critical overdensity for spherical collapse and $\sigma$ is the variance of the perturbations on the scale $R$ corresponding to the mass $M$, viz.
\begin{equation}
\sigma^2[R(M)] = \frac{1}{(2 \pi)^3} \int{P_{\delta}(k) |W(kR)|^2 d^3k},
\label{eq: sigmavardef}
\end{equation}
where $W(kR)$ is the Fourier transform of the spherical top hat function and the density power spectrum $P_{\delta}(k, z)$.

To compute the mass function through Eq.~(\ref{eq: mf}), one has to choose an expression for $\nu \varphi(\nu)$. We adopt the \citet{Sheth:1999su} function
\begin{equation}
\nu \varphi(\nu) = A \sqrt{\frac{2 a \nu^2}{\pi}} [1 + (a \nu^2)^{-p}] \exp{(-a \nu^2/2)},
\label{eq: stmf}
\end{equation}
with $(A, a, p) = (0.322, 0.75, 0.3)$. Although many other choices are possible (see for instance the extensive list in \cite{2013A&C.....3...23M} and refs therein), we note that the choice of the mass function is not critical for our aims, since we are mainly interested in comparing the impact of different MC relations on peak count rather than forecasting exact numbers.

Not all the clusters will give rise to detectable peaks, but only those with a $S/N$ ratio larger than a fixed threshold. We have first to take into account that the shot noise from discrete background galaxy positions and the intrinsic ellipticity distribution introduce a scatter of the observed aperture mass $M_{\rm ap}$ around its theoretically expected value $\hat{M}_{\rm ap}(M)$. As a consequence, a halo of mass $M$ has a certain probability $p(M_{\rm ap} | M)$ to produce an aperture mass $M_{\rm ap}$ which we can model as a Gaussian, namely
\begin{equation}
p(M_{\rm ap} | M) \propto \exp{\left \{ - \frac{1}{2} \left [ \frac{M_{\rm ap} - \hat{M}_{\rm ap}(M)}{\sigma_{\rm ap}} \right ]^2 \right \}} \ .
\label{eq: mapprob}
\end{equation}
The probability that the $S/N$ ratio will be larger than a given threshold will read \citep{Bartelmann:2002dh}
\begin{equation}
p({\cal{S}} > {\cal{S}}_{\rm th} | M_{\rm vir}, z) = \frac{1}{2} {\rm erfc}\left [ \frac{{\cal{S}}(M_{\rm vir}, z) - {\cal{S}}_{\rm th}}{\sqrt{2}} \right ].
\label{eq: snrprob}
\end{equation}
Therefore, the number density of haloes giving a detectable weak lensing peak will be the product of the halo mass function and this probability, i.e.
\begin{equation}
{\cal{N}}_{\rm lens}(M, z) = p({\cal{S}} > {\cal{S}}_{\rm th} | M, z) {\cal{N}}(M, z) \ .
\label{eq: nlens}
\end{equation}
Integrating over the redshift and the mass and multiplying by the survey area gives the total number of peaks generated by cluster haloes and with $S/N$ larger than a threshold value ${\cal{S}}_{\rm th}$, which reads
\begin{eqnarray}
N_{\rm halo}({\cal{S}} > {\cal{S}}_{\rm th}) & = & \left ( \frac{c}{H_0} \right )^3 \left ( \frac{\pi}{180} \right )^2 \left ( \frac{\Omega}{1 \ {\rm deg}^2} \right ) \\
~ & \times &  \int_{z_{L}}^{z_{U}}{\frac{r^2(z)}{E(z)} dz \nonumber
\int_{0}^{\infty}{{\cal{N}}_{\rm lens}(M, z) dM}},
\label{eq: npktot}
\end{eqnarray}
with $r(z) = (c/H_0)^{-1} \chi(z)$. As redshift limits, we set $(z_{L}, z_{U}) = (0.1, 1.4)$ since the number of peaks outside this range is negligible---although the survey will likely detect galaxies over a much larger range.

The number of observed peaks is the sum of ${\cal{N}}_{\rm halo}$ and a term due to the contamination from the LSS,
\begin{equation}
{\cal{N}}_{\rm pk}({\cal{S}} > {\cal{S}}_{\rm th}) = {\cal{N}}_{\rm halo}({\cal{S}} > {\cal{S}}_{\rm th}) +{\cal{N}}_{\rm LSS}({\cal{S}}_{\rm th}),
\end{equation}
where the LSS term reads \citep{2010A&A...519A..23M,Maturi:2011am}
\begin{equation}
{\cal{N}}_{\rm LSS} = \frac{1}{(2 \pi)^{3/2}} \left ( \frac{\sigma_{\rm LSS}}{\sigma_{\rm ap}} \right )^2
\frac{\kappa_{\rm th}}{\sigma_{\rm ap}} \exp{\left [ - \frac{1}{2} \left ( \frac{\kappa_{\rm th}}{\sigma_{\rm ap}} \right )^2 \right ]},
\label{eq: npklss}
\end{equation}
with $\kappa_{\rm th} = {\cal{S}}_{\rm th} \sigma_{\rm ap}$ and
\begin{equation}
\left ( \frac{\sigma_{\rm LSS}}{\sigma_{\rm ap}} \right )^2 = \frac{\int_{0}^{\infty}{P_N(\ell) \left | \hat{\Psi}(\ell) \right |^2 \ell^3 d\ell}}
{\int_{0}^{\infty}{P_{\varepsilon}(\ell) \left | \hat{\Psi}(\ell) \right |^2 \ell d\ell}} \ .
\label{eq: defsigmaratio}
\end{equation}
${\cal{N}}_{\rm LSS}$ only depends on the noise properties and the threshold $S/N$ ratio, but not on the lens mass and redshift. This is an obvious consequence of this term being due to the LSS rather than a particular cluster. For this same reason, ${\cal{N}}_{\rm LSS}$ is determined by the matter power spectrum (and hence the underlying cosmological scenario) entering $P_{N}(\ell)$.

\subsection{Cumulative peak number and $S/N$ threshold}

As a preliminary step, it is worth investigating how the total number of peaks (actual ones due to clusters and fake ones due to LSS) change as a function of the threshold $S/N$ ratio. This will also tell us how to choose the threshold $S/N$ value to discriminate between true and fake peaks. 

Fig.~\ref{fig: npktot} helps us to highlight some important issues. First, in the left panel, we plot the ${\cal{N}}_{\rm pk}({\cal{S}} > {\cal{S}}_{\rm th})$ as a function of the threshold $S/N$ ratio for the six different MC relations in Table~1. Somewhat surprisingly, although the filter has been set using the D08 relation as fiducial, the number of peaks is larger for the other relations. This is better shown in the central panel where the number of peaks is scaled with respect to the D08 one. Note that the quick increase of the ratio for large ${\cal{S}}_{\rm th}$ is not due to the number of peaks diverging, but rather to ${\cal{N}}_{\rm pk}({\cal{S}} > {\cal{S}}_{\rm th})$ quickly approaching the null value for the reference D08 case.

\begin{figure*}
\resizebox{\hsize}{!}
{\includegraphics[width=7.0cm]{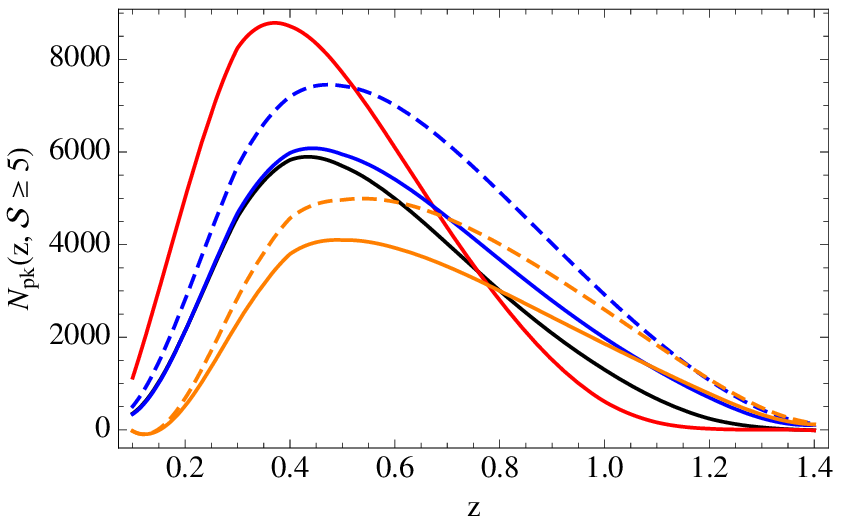}
\includegraphics[width=7.0cm]{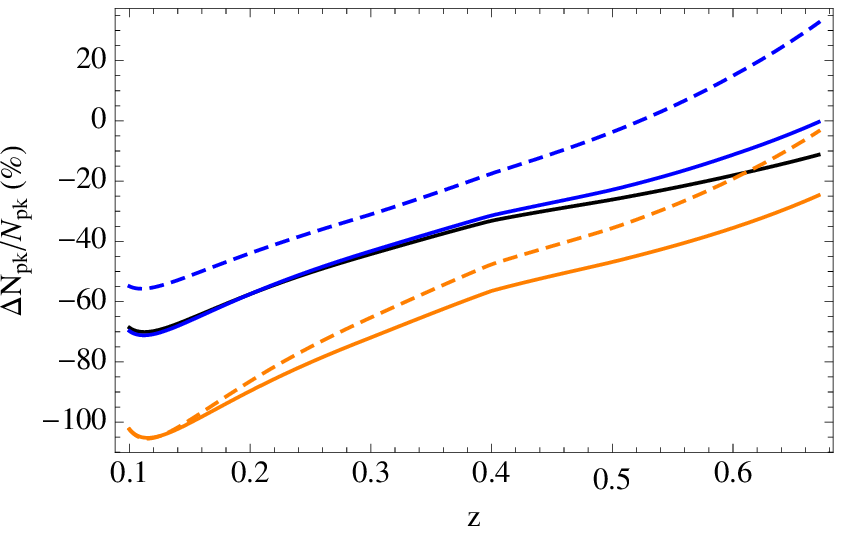}}
\caption{{\it Left.} Number of peaks with ${\cal{S}} \ge 5$ in redshift bins of width $\Delta z = 0.1$ as a function of the bin redshift. {\it Right.} Percentage deviation of the number of peaks per bin with respect to D08 (only reporting the range where ${\cal{N}}_{\rm pk}(z,{\cal{S}}  \ge 5)$ for D08 is significantly non vanishing). Black, red, blue, dashed blue, orange and dashed orange lines refer to B01, D08, Ok10, Ok10z, Og12 and Og12z, respectively.}
\label{fig: npkzed}
\end{figure*}

The larger number of peaks for MC relations others than the fiducial D08 one can be traced back to the higher $S/N$ values for clusters of mass $\log{M_{200}} > 15$, where, hereafter, $M_{200}$ is the mass in solar units. The higher the cluster redshift, the larger the mass to pass the selection threshod. As $z$ increases, the minimum mass a cluster should have to generate a detectable peak increases too, but its value also depends on the adopted MC relation. Since the D08 model predicts the smaller $S/N$ values, the limiting mass is larger for this case so that the contribution to the total number of peaks in the highest redshift bins is smaller and smaller as the threshold $S/N$ increases. As a consequence, the MC relations predicting larger ${\cal{S}}(z, M)$ have a larger chance to produce haloes massive enough to pass the selection threshold thus leading to the greater ${\cal{N}}_{\rm pk}$ values.

Up to now, we have considered the total number of detectable peaks, but what is actually of interest to investigate the MC relation is the number of peaks due to clusters only. The right panel in Fig.~\ref{fig: npktot}, showing the ratio between the number of peaks due to clusters and the total ones, helps us to disentangle the clusters from fake peaks. As expected, ${\cal{N}}_{\rm pk}$ is dominated by the LSS term for small $S/N$ ratio, that is to say, the smaller is the $S/N$ ratio, the larger is the probability that the detected peak is a fake one due to the LSS rather than the evidence for a cluster. This is in agreement with common sense expectation and previous analysis in literature using different cosmological models and survey parameters \citep{2005A&A...442...43H,2010A&A...519A..23M}. The ratio ${\cal{N}}_{\rm lens}/{\cal{N}}_{\rm pk}$ is, however, a strong increasing function of ${\cal{S}}_{\rm th}$ for ${\cal{S}}_{\rm th}>2.5$. Imposing the requirement ${\cal{N}}_{\rm lens}/{\cal{N}}_{\rm pk} > 0.
9$, one gets ${\cal{S}}_{\rm th} \simeq 5$ with a very weak dependence on the adopted MC relation. We can therefore safely argue that all the peaks with ${\cal{S}} \ge 5$ are due to clusters. Note that we find a value for ${\cal{S}}_{\rm th}$ comparable but larger than what is suggested in \cite{2010A&A...519A..23M} because of differences in both the cosmological model and the survey characteristics.

\subsection{Number of peaks in redshift bins}

The total number of peaks is obtained by integrating over the full redshift range as in Eq.~\eqref{eq: npktot}. This obviously degrades the information on the dependence of the MC relation on $z$. It is worth investigating what can be learned by binning the peaks according to their redshift. The number of peaks in a bin centred on $z$ and with width $\Delta z$ can be computed using again Eq.~(\ref{eq: npktot}) and replacing $(z_{L}, z_{U})$ with $(z - \Delta z/2, z + \Delta z/2)$. Since we need a redshift measurement to assign a given peak to a bin, we implicitly assume that all the detected peaks are due to clusters, i.e. ${\cal{N}}_{\rm pk}(z) = {\cal{N}}_{\rm halo}(z)$; in other words, we disregard the LSS term. Actually, for a given threshold $S/N$ ratio, such a term provide a constant contribute to the number of peaks in each redshift bin. However, since we will only consider peaks with ${\cal{S}}_{\rm th} > 5$, one can be confident that the predicted numbers indeed refers to peaks with measurable redshift. For $\Delta z = 0.1$, we get the ${\cal{N}}_{\rm pk}(z, {\cal{S}} > {\cal{S}}_{\rm th})$ curves shown in the left panel of Fig.~\ref{fig: npkzed} for the different MC relations listed in Table~\ref{tab: mclist}.

Binning the data indeed helps to better discriminate among the different MC relations. The number of peaks in each given bin happens to be quite different from one MC relation to another---with D08 dominating the signal for $z < 0.5$, but quickly decreasing for larger $z$. Indeed, a way to discriminate between D08 and other relations is by looking at bins with $z > 1$, where almost no peaks are expected for the D08 case, while a still significant number can be found for other relations. In particular, Ok10z and Og12z give the largest ${\cal{N}}_{\rm pk}(z)$ values.

That redshift binning improves the efficiency in discriminating among different MC relations can also be quantitatively shown by considering firstly the total number of clusters given in Table~\ref{tab: npklist}. The comparison of numbers for the relative difference with respect to the D08 case (reported in the third column\footnote{It is worth noting that the difference in the total number of peaks and the difference in the binned peaks can have a different sign depending on which redshift bin is considered. Indeed, in the first case, we are referring to the total area under the curve in the left panel of Fig.\,4, while, in the second case, we consider the area under only a portion of the curve. For instance, since the B01 line stays almost always under the D08 one, it is clear that the $\Delta$ in Table\,2 takes a negative value. Nonetheless, if we refer only to the peaks in a bin with $z > 1$, the D08 line is lower than the B01 one so that the difference is now positive.}) with the ones referred to binned 
data (which can be read from the right panel of Fig.~\ref{fig: npkzed}) convincingly shows that binning in $z$ is utterly effective. 

\begin{table}
\centering
\caption{\label{tab: npklist}Total number of peaks ${\cal{N}}_{\rm pk}({\cal{S}} \ge 5)$ and percentage deviation $\Delta = \left [
{\cal{N}}_{\rm pk}^{id}({\cal{S}} \ge 5) - {\cal{N}}_{\rm pk}^{D08}({\cal{S}} \ge 5) \right ]/{\cal{N}}_{\rm pk}^{D08}({\cal{S}} \ge 5)$ for the different MC relations considered.}
\begin{tabular}{ccc}
\hline
Id & ${\cal{N}}_{\rm pk}({\cal{S}} \ge 5)$ & $\Delta \  (\%)$ \\
\hline \hline\\
B01 & 17506 & -25 \\
D08 & 23262 & ---- \\
Ok10 & 19922 & -14 \\
Ok10z & 26234 & +13 \\
Og12 & 13957 & -40  \\
Og12z & 17981 & -23 \\
\hline
\end{tabular}
\end{table}

\section{Fisher matrix forecasts}
In order to quantify the conclusions discussed above, we carry on a Fisher matrix analysis and consider as observed data the total number of peaks with ${\cal{S}} > {\cal{S}}_{\rm th}$ in equally spaced redshift bins centred on $z$ and with width $\Delta z = 0.1$ over the range $(0.1, 1.4)$. As usual when dealing with number counts, we can assume Poisson errors and then quantify the agreement between data and model through the following likelihood function \citep{Cash:1979vz}
\begin{equation}
-2 \ln{{\cal{L}}}({\bf p}) = - 2 \sum_{i = 1}^{{\cal{N}}_{bin}}{\nu_i \ln{\lambda_i} - \lambda_i - \ln{\nu_i !}}
\label{eq: deflike}
\end{equation}
where, to simplify the notation, we have respectively defined $\lambda_i = {\cal{N}}_{\rm pk}^{\rm th}(z_i, {\bf p})$ and $\nu_i = {\cal{N}}_{\rm pk}^{obs}(z_i)$ for the theoretical and observed number of peaks in the $i$th redshift bin, ${\bf p}$ denotes the set of parameters we want to constrain and the sum runs over the ${\cal{N}}_{bin}$ bins. The Fisher matrix elements will be given by the second derivatives of the logarithm of the likelihood with respect to the parameters of interest evaluated at the fiducial values. Starting from Eq.~\eqref{eq: deflike}, one gets \citep{Holder:2001db}
\begin{equation}
F_{ij} = - \frac{\partial^2 \ln{{\cal{L}}}}{\partial p_i \partial p_j} =
\sum_{k = 1}^{{\cal{N}}_{bin}}{\frac{\partial \lambda_k}{\partial p_i} \frac{\partial \lambda_k}{\partial p_j} \frac{1}{\lambda_{k}^{\rm fid}}}
\label{eq: fij}
\end{equation}
where $\lambda_{k}^{\rm fid}$ is the expected number of peaks in the $k$th bin for the fiducial model. The covariance matrix is then simply the inverse of the Fisher matrix and its diagonal elements represent the lowest variance one can achieve on the model parameter measurement.

Although our main interest focusses on the MC relation, peak number counts do not depend on this relation only. On the contrary, ${\cal{N}}_{\rm pk}(z)$ strongly depends on the background cosmological model too so that the Fisher matrix must be computed with respect to both sets of parameters. As a first approximation, one can hold the cosmology fixed and only derive the constraints on the MC parameters.\footnote{To this end, one has simply to remove the corresponding rows and columns from the Fisher matrix, while marginalisation can be obtained by deleting them from the covariance matrix and then inverting back to get the marginalised Fisher matrix.} Otherwise, one can assume the MC relation known from a different probe and investigate to which extent peak number counts can constrain cosmology. Both possibilities will be considered below.

\subsection{Constraints on the MC parameters}

\begin{figure}
\centering
\includegraphics[width=0.5\textwidth]{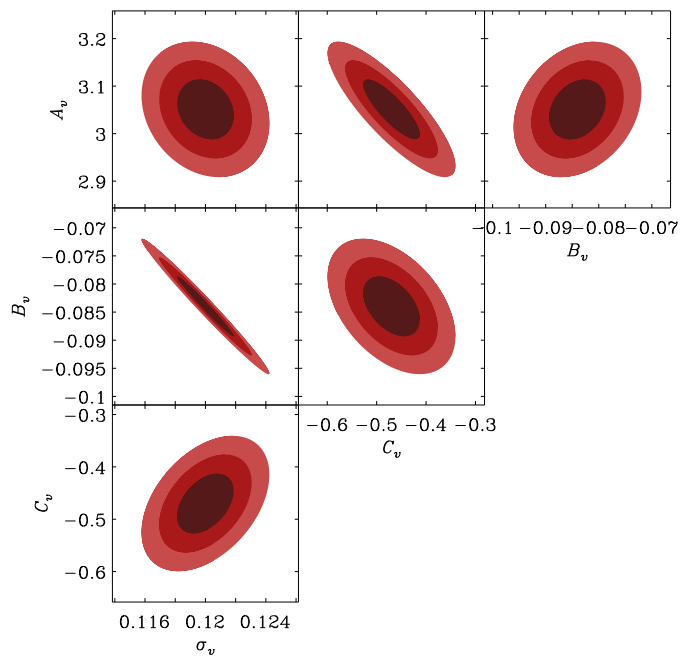}
\caption{Fisher matrix forecasts for the $68, 95, 99\%$ CL assuming a fiducial D08 MC relation and a filter aperture $\vartheta = 2 $\,arcmin.}
\label{fig: fmplots}
\end{figure}

Let us first consider the case with the background cosmological model held fixed. An implicit assumption which the Fisher matrix forecast relies on is that the confidence regions may be approximated as a Gaussian ellipsoids, while it is not uncommon that the true ones have broad tails or significant curvature. \citet{Holder:2001db} investigated whether this is the case for number counts by comparing with Monte Carlo analysis of simulated datasets. They found that Fisher matrix forecasts can indeed be trusted. Therefore, we are confident that our estimated iso\,-\,likelihood contours, shown in Fig.~\ref{fig: fmplots}, provide reliable accounts of the degeneracies in the MC relation parameters space. It is worth emphasising that these results have been obtained assuming that the cosmological parameters are known with infinite precision so that they can be hold fixed in the Fisher matrix derivation. We will return later to this point.

The marginalised $1\sigma$ errors on the MC parameters for the fiducial are
\begin{equation}
\sigma(A_v) = 0.04 \ , \ \sigma(B_v) = 0.003 \ , \ \sigma(C_v) = 0.04 \ , \ \sigma(\sigma_v) = 0.001,
\end{equation}
which can also conveniently be rewritten as
\begin{equation}
\Delta(A _v)= 1\% \ , \
\Delta(B _v) = 4\%  \ , \
\Delta(C _v) = 9\% \ , \
\Delta(\sigma _v) = 1\% ,
\end{equation}
with $\Delta(p) = \sigma(p)/p$. Such numbers nicely show that peak number counts in redshift bins provide competitive constraints on the MC parameters thus enabling us to discriminate convincingly among different MC relations. Indeed, comparing the differences of the $(A_v, B_v, C_v)$ values in Table~\ref{tab: mclist} with the $1\sigma$ uncertainties given above, we can safely conclude that the B01, Ok10 and Og12 relations could be rejected with high confidence, should the actual MC relation coincide with fiducial D08.

It is worth wondering whether the results depend on the adopted fiducial MC relation. We do not expect this to be the case since the Fisher matrix approach should provide a reliable description of the likelihood in the neighbourhood of the fiducial values whatever these values are. However, the Ok10 and Og12 mass slope parameters are so far away from those of D08 that some failure of the Fisher matrix can not be excluded a priori. In order to check this, we have therefore repeated the Fisher matrix evaluation taking the Og12z as fiducial model for the MC relation while holding the cosmological parameters set to the Planck ones. We get\,:

\begin{equation}
\sigma(A_v) = 1.1 \ , \ \sigma(B_v) = 0.04 \ , \ \sigma(C_v) = 0.4 \ , \ \sigma(\sigma_v) = 0.02 \ ,
\end{equation}
for the marginalised $1\sigma$ errors, i.e.
\begin{equation}
\Delta(A _v)= 14\% \ , \
\Delta(B _v) = 6\%  \ , \
\Delta(C _v) = 80\% \ , \
\Delta(\sigma _v) = 13\% \ .
\end{equation}
Although there is a significative degradation of the constraints due to the degeneracy\footnote{Such a degeneracy is partly due to a numerical coincidence related to the choice of the pivot mass and the similarity of the $(B_v, C_v)$ values. Indeed, over the mass range probed, the quantity $(M_{200}/M_{piv})^{B_v}$ takes values close to $(1 + z)^{C_v}$ so that the two terms can hardly be distinguished.} between $B_v$ and $C_v$, it is nevertheless still possible to discriminate among the different MC relations. Using the value of the mass slope parameter $B_v$ as discriminator, we now get $|B_v(Og12z) - B_v(mc)|/\sigma(B_v) = (11.5, 12.6, 4.8)$ for $mc = $ B01, D08,  Ok10 so that it is still possible to discriminate among D08 and empirically motivated MC relations in agreement with the previous result.

The above constraints have been obtained assuming that the cosmological parameters are perfectly known. Relaxing this assumption introduces degeneracies which significantly enlarge the confidence ranges. For a cosmological model with DE EoS described by the CPL ansatz and fitting both the MC and seven cosmological parameters $(\Omega_M, \Omega_b, w_0, w_a, h, n_{PS}, \sigma_8)$, the marginalised $1\sigma$ errors on the MC parameters now read
\begin{equation}
\sigma(A_v) = 0.80 \ , \ \sigma(B_v) = 0.05 \ , \ \sigma(C_v) = 1.2 \ , \ \sigma(\sigma_v) = 0.04 .
\end{equation}
It is possible to discriminate among empirically (e.g. Ok10, Og12) and numerically inspired MC relations thanks to the radically different $B_v$ value.

The situation can be improved by adopting an intermediate strategy. Forcing the model to have a $\Lambda$ term (i.e. setting $w_0 = -1$ and $w_a = 0$) and fixing $(\Omega_b, h, n_{PS})$ to their fiducial values, we get the following $1 \sigma$ errors
\begin{equation}
\sigma(A_v) = 0.06 \ , \ \sigma(B_v) = 0.007 \ , \ \sigma(C_v) = 0.05 \ , \ \sigma(\sigma_v) = 0.02.
\end{equation}
Albeit the constraints are larger by roughly a factor of two than in the case obtained fixing background cosmology (but note that this is not the case for $C_v$), they are still remarkably strong allowing for discrimination among different MC relations.

Although our focus here is on the use of peak number counts, this is not the only probe one can use. Degeneracies among cosmological and MC parameter can indeed be lifted by using different tracers. To this end, we combine our Fisher matrix with the one obtained by inverting the Planck covariance matrix\footnote{We use the covariance matrix corresponding to the joint fit of Planck and WMAP polarisation data. Note that (after marginalising over nuisance parameters) this constrains $(\Omega_M h^2, \Omega_b h^2, \Theta, n_{PS}, \ln{ {\cal{A}}_s})$ with $\Theta$ the angular scale of the sound horizon and ${\cal{A}}_s$ the power spectrum normalisation. We therefore first invert the covariance matrix and then project the corresponding Fisher matrix on our seven dimensional parameter space $(\Omega_M, \Omega_b, w_0, w_a, h, n_{PS}, \sigma_8)$.}. By letting all the seven cosmological and four MC relation parameters free to change, we get

\begin{equation}
\sigma(A_v) = 0.13 \ , \ \sigma(B_v) = 0.007 \ , \ \sigma(C_v) = 0.07 \ , \ \sigma(\sigma_v) = 0.003,
\end{equation}
which are dramatically smaller than the case with no Planck data and comparable (although larger) with those for the case with the cosmology set to the fiducial case. Setting $(\Omega_b, w_0, w_a, h, n_{PS})$ to their fiducial values does improve now only the constraint on $A_v$ (from 0.13 to 0.06), while those on $(B_v, C_v, \sigma_v)$ are almost left unchanged. This is expected since the Planck data already set strong limits on the background cosmological model so that forcing it to be equal to the fiducial one has now a minor impact on the MC parameters confidence ranges.

\subsection{Constraints on cosmological parameters}
\begin{figure}
\centering
\includegraphics[width=0.45\textwidth]{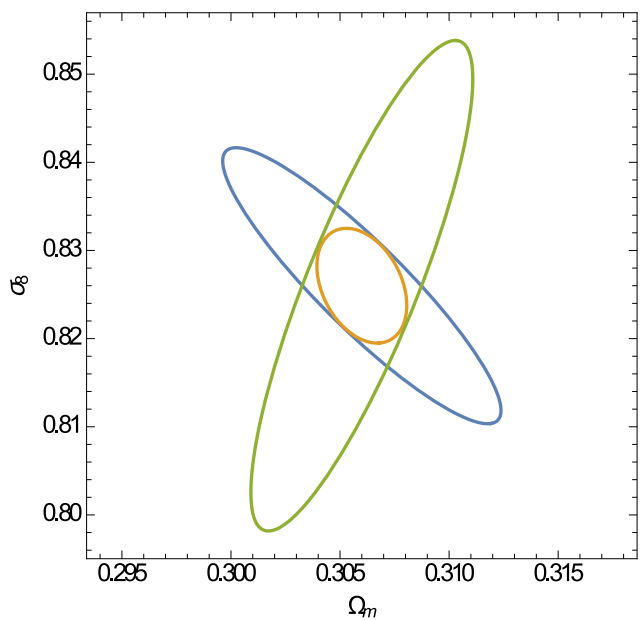}
\caption{Marginal error $1\sigma$ contours in the $(\Omega_M, \sigma_8)$ plane are shown as obtained from peak count alone (blue ellipse), Planck alone (green ellipse) and the combination of the two probes (yellow ellipse).}
\label{fig:ellipses_case3}
\end{figure}

Let us now explore the use of peak number counts as a tool to probe the background cosmological model. The most favourable case is that with the smallest number of parameters so that we only vary $(\Omega_M, \sigma_8)$ and set all the remaining cosmological and MC quantities to their fiducial values. Using peaks only, we get
\begin{equation}
\sigma(\Omega_M) = 0.004 \ , \ \sigma(\sigma_8) = 0.01 ,
\end{equation}
which are already comparable to what one can obtain using the Planck data alone. A joint fit to peak number counts and Planck further pushes down the errors leading to
\begin{equation}
\sigma(\Omega_M) = 0.001 \ , \ \sigma(\sigma_8) = 0.004.
\end{equation}
This can be easily understood by looking at Fig.~\ref{fig:ellipses_case3}, where $1\sigma$ marginal error contours in the $(\Omega_M, \sigma_8)$ plane are shown as obtained from peak count alone (blue ellipse), Planck alone (green ellipse) and the combination of the two probes (yellow ellipse). As a matter of fact, weak lensing and CMB temperature anisotropies suffer from different degeneracy for what concercs those two parameters, and this happens in a way that makes their combination more effective.

\begin{figure*}
\centering
\includegraphics[width=0.25\textwidth]{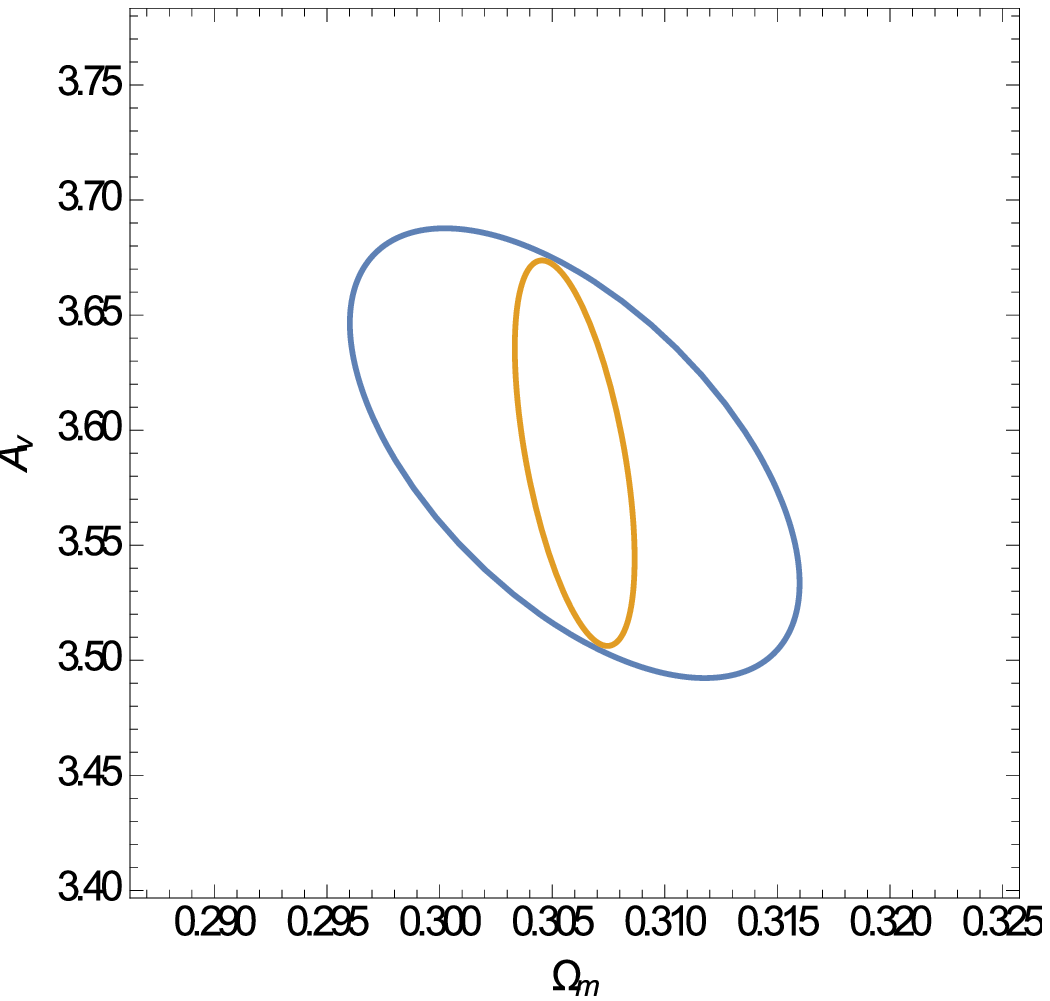}\includegraphics[width=0.25\textwidth]{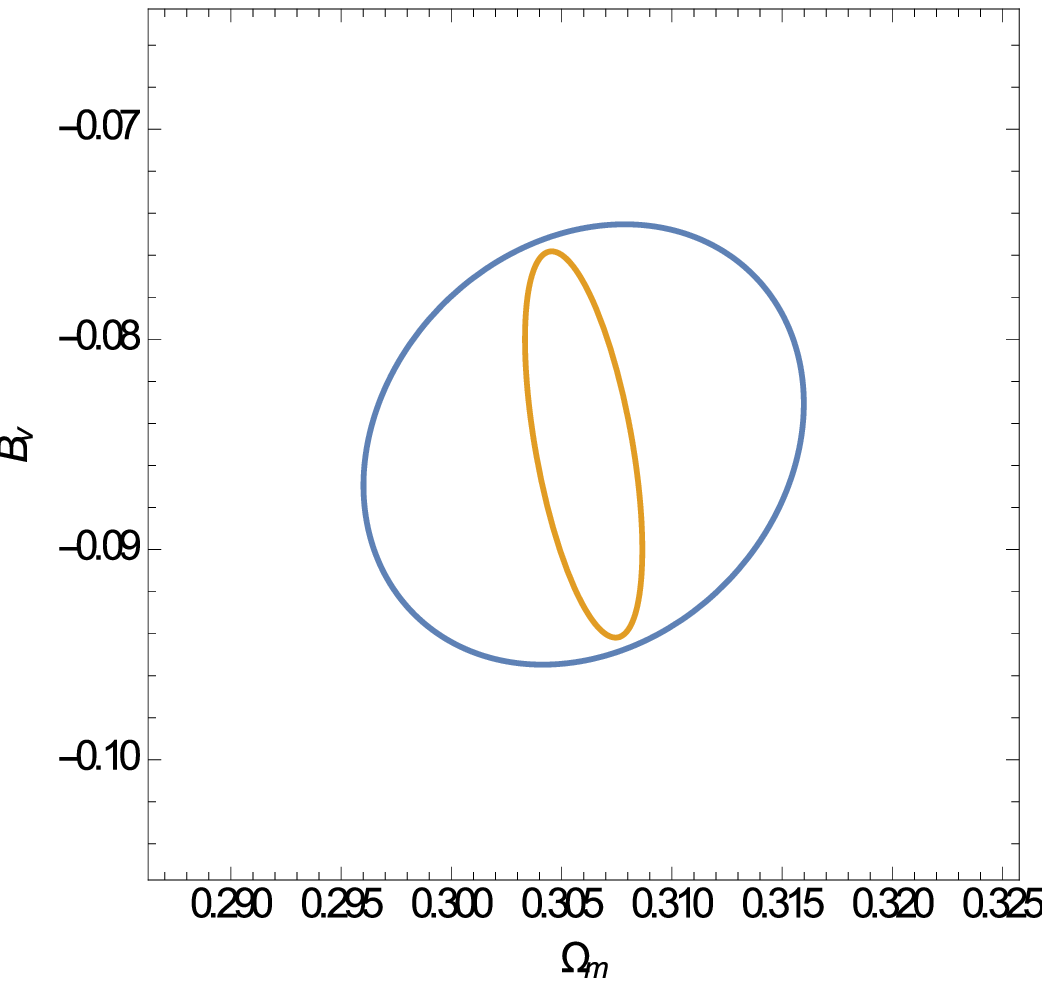}\includegraphics[width=0.25\textwidth]{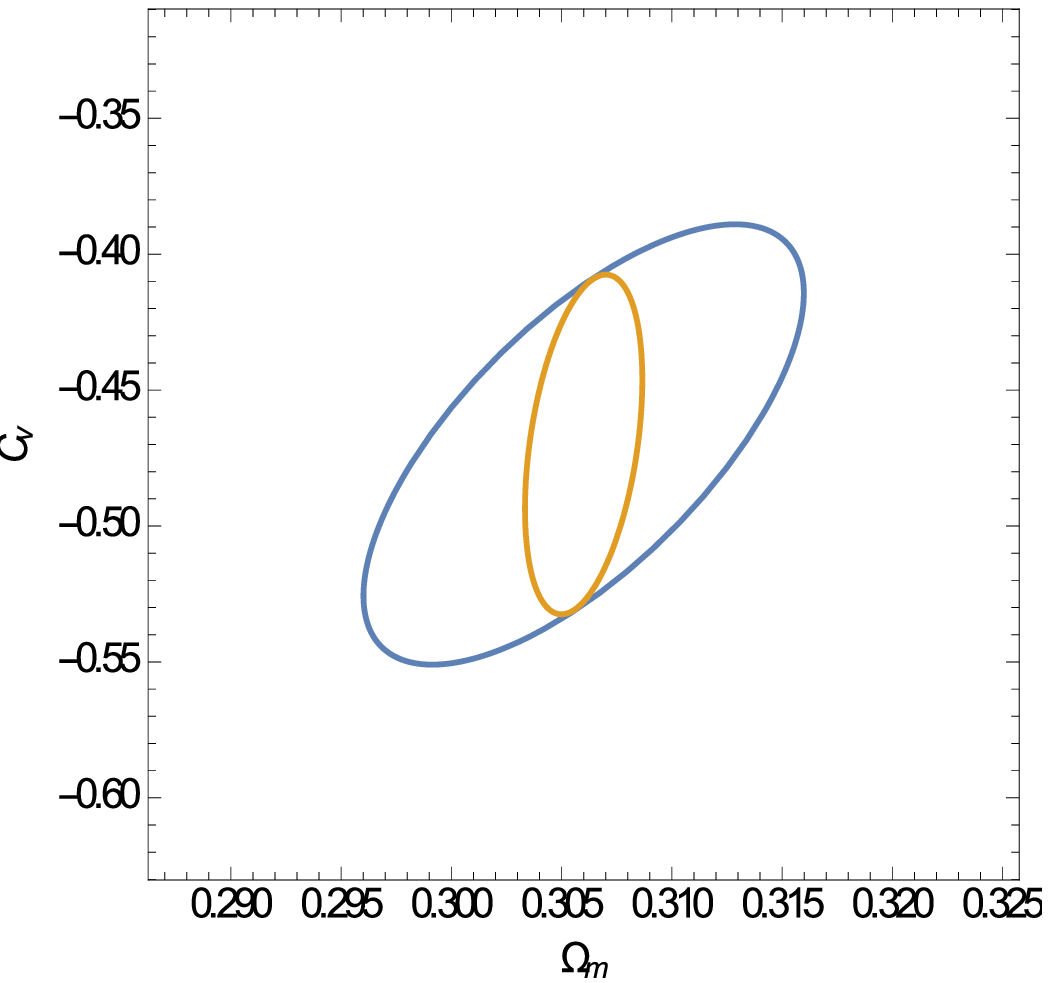}\includegraphics[width=0.25\textwidth]{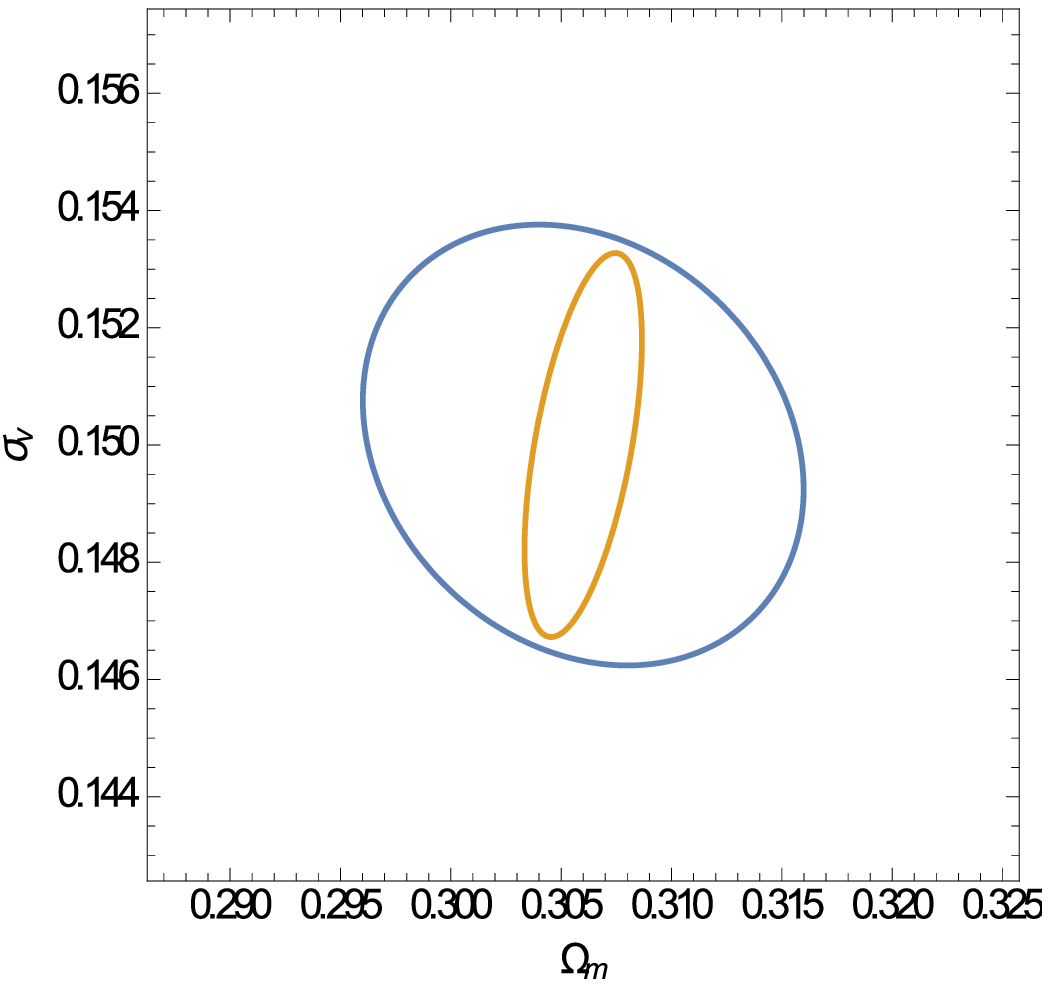}\\\includegraphics[width=0.25\textwidth]{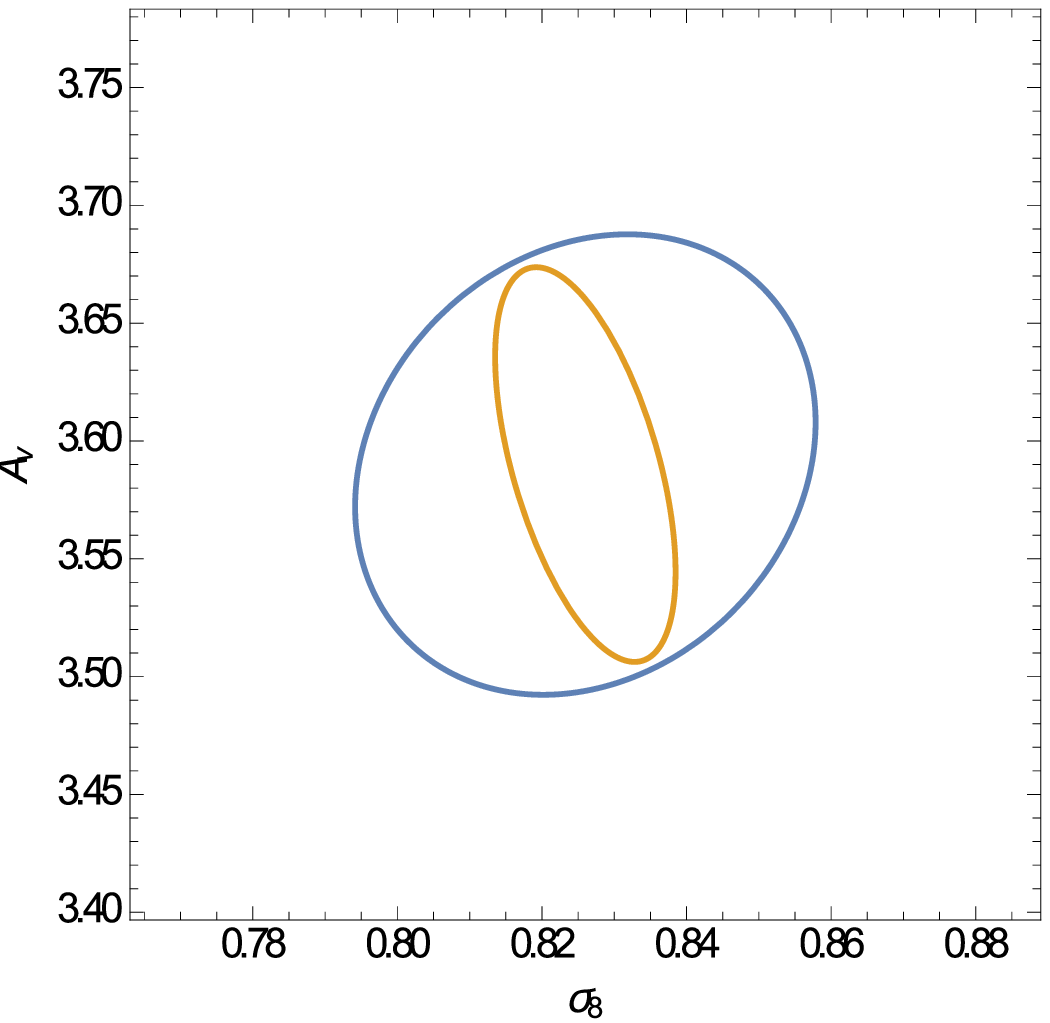}\includegraphics[width=0.25\textwidth]{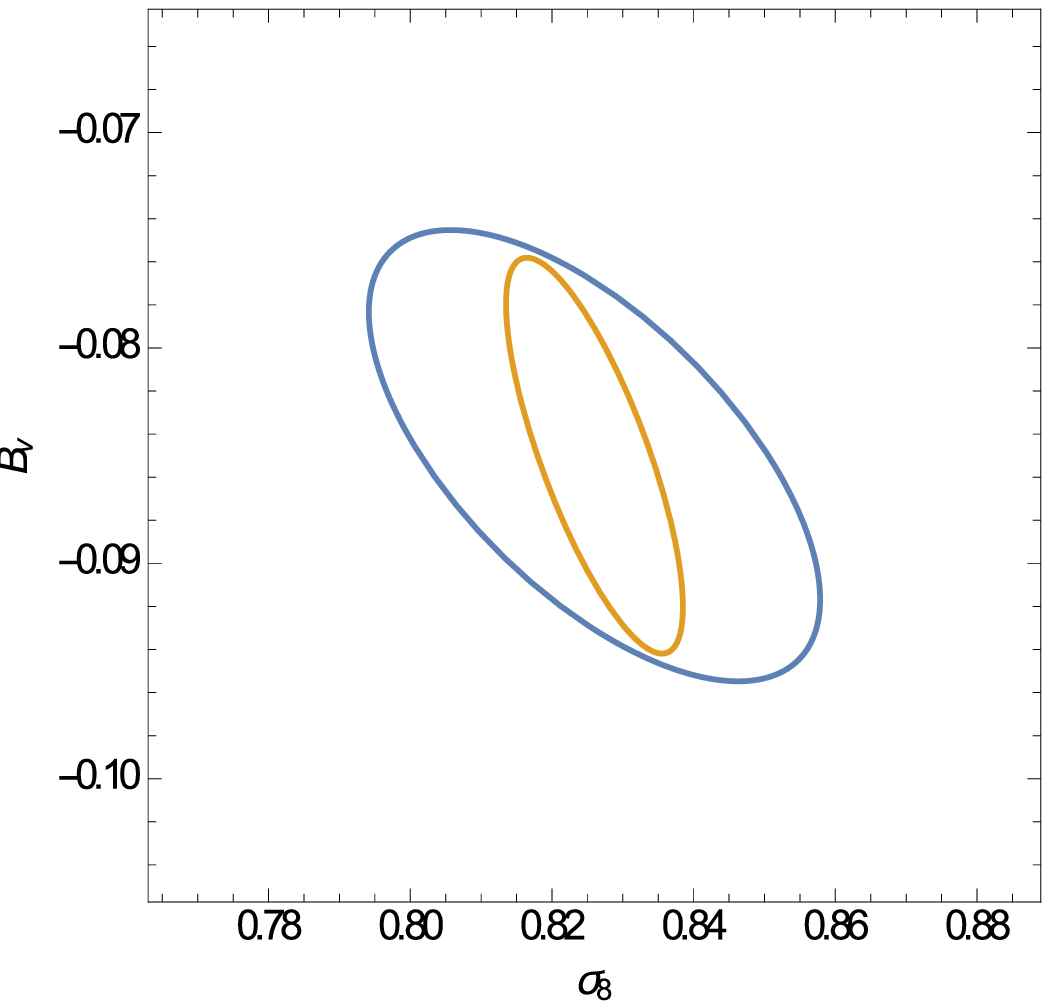}\includegraphics[width=0.25\textwidth]{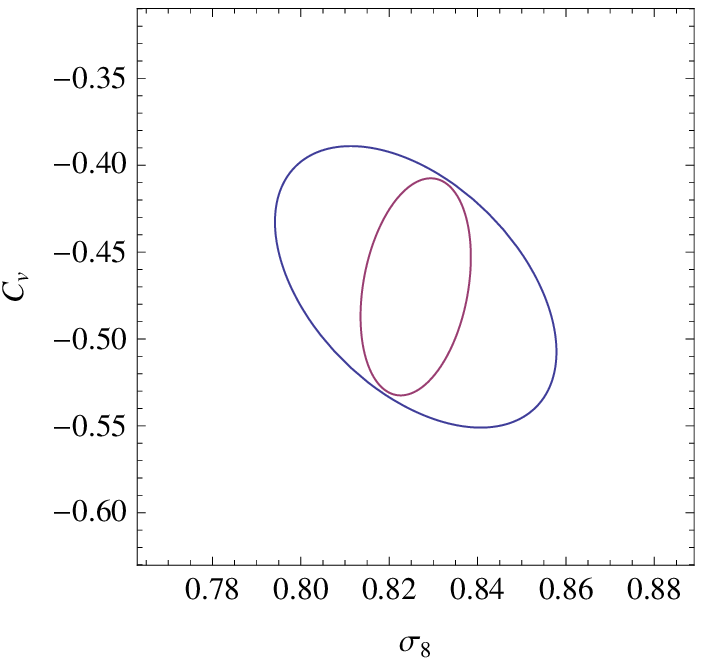}\includegraphics[width=0.25\textwidth]{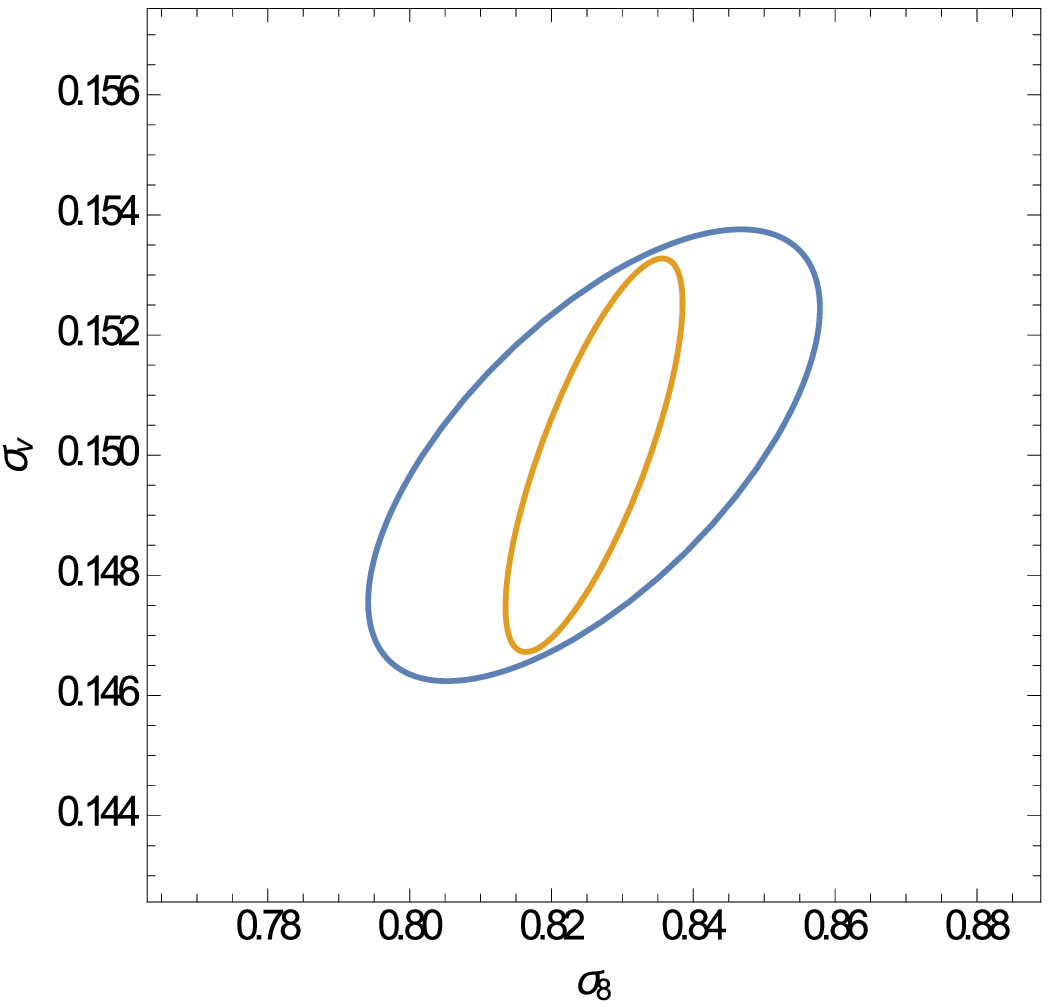}
\caption{Marginal error $1\sigma$ contours for all the combinations of MC parameters with $\Omega_m$ (top panels) and $\sigma_8$ (bottom panels), for the case with (yellow curves) or without (blue curves) Planck priors.}
\label{fig:ellipses_case4}
\end{figure*}

Moving towards a more realistic case, we can allow the MC parameters to change hence introducing degeneracies among $(\Omega_M, \sigma_8)$ and $(A_v, B_v, C_v, \sigma_v)$. Hence, we now find
\begin{equation}
\sigma(\Omega_M) = 0.006 \ , \ \sigma(\sigma_8) = 0.02
\end{equation}
from peak number counts only. Instead, if we add Planck we have
\begin{equation}
\sigma(\Omega_M) = 0.001 \ , \ \sigma(\sigma_8) = 0.008.
\end{equation}
Although the constraints have been weakened, the uncertainties on the MC parameters have not dramatically broadened the confidence ranges of $(\Omega_M, \sigma_8)$, which are still well constrained even when only peak number counts are used. Fig.~\ref{fig:ellipses_case4} illustrates this by plotting the marginal error ellipses for all the combinations of MC parameters with $\Omega_m$ and $\sigma_8$, for the case with (yellow curves) or without (blue curves) Planck priors.

As a further step, we investigate the precision of peaks number counts to constrain all cosmological parameters under the assumption that the MC relation has been already constrained through a different method. Here, we only consider the case when Planck data are added to peaks since the degeneracies among cosmological parameters can not be lifted by a probe only sensible to integrated quantities. Adding Planck and peak Fisher matrices and marginalising over the MC parameters gives
\begin{equation}
\sigma(\Omega_M) = 0.006 \ , \ \sigma(\Omega_b) = 0.001 \ , \ \sigma(w_0) = 0.09 \ , \ \sigma(w_a) = 0.51
\end{equation}
and
\begin{equation}
\sigma(h) = 0.009 \ , \ \sigma(n_{PS}) = 0.006 \ , \ \sigma(\sigma_8) = 0.02 ,
\end{equation}
which nicely compares to what can be obtained combining Planck with other data such as BAOs and SNeIa. In particular, there is a significant improvement in the constraints on the dark energy equation of state. Specifically, we get
\begin{equation}
\Delta(w_0) = 9\% \ , \ \Delta(1 + w_a) = 51\% ,
\end{equation}
to be compared with $18\%$ and $51\%$ from the combination of Planck, BAOs and SNeIa data. However, it is worth stressing that the comparison is actually unfair given that we have here contrasted future peaks statistics with present day SNeIa and BAO data. A fair comparison would ask for a preliminary Fisher matrix forecast based on, e.g., Euclid BAO data which is, however, outside our aims here.

Moroever, let us emphasise that this result is strongly bound to the assumption that the MC relation is perfectly known. If we relax it, the constraints in the eleven-dimensional parameter space strongly weaken. While $(\Omega_M, \Omega_b, h, n_{PS}, \sigma_8)$ are still reasonably well constrained, the accuracy on dark energy parameters $(w_0, w_a)$ read
\begin{equation}
\Delta(w_0) = 41\% \ , \ \Delta(1 + w_a) = 131\% ,
\end{equation}
asking for further data to narrow down the confidence ranges.

\subsection{The impact of baryons}
Both the B01 and D08 relations and the Sheth\,-\,Tormen mass function have been inferred from the results on $N$-body simulations only including collisionless dark matter particles. As a matter of fact, galaxy clusters also contain baryons (both in galaxies and in hot gas), so that a realistic description should take their presence into account. As a consequence, we should also investigate the impact of baryons on peak number counts and hence on the constraints discussed above. While addressing how the MC and mass function are changed by the collapse of baryons is outside our aims, we can nevertheless draw some lessons considering recent results on this issue.

First, we note that baryons can alter the halo concentration thus changing the normalisation of the MC relation. However, we expect that this effect is less and less important as the halo mass increases. Indeed, Fig.~8 in Duffy et al. (2010) shows that the ratio $c_{\rm vir}^{bar}/c_{\rm vir}^{DM}$ (with $c_{\rm vir}^{bar}$ and $c_{\rm vir}^{DM}$ the halo concentration with and without baryons) deviates less than $10\%$ from unity for $\log{M_{\rm vir}} > 13.5$, as inferred comparing CDM only with hydrodynamical simulations including baryons. Although this is not a direct evidence that the MC relation is unaffected, we can reasonably infer that the mass and redshift power\,-\,law dependence we have adopted so far is a good approximation even when baryons are included. It is worth stressing that, while it is possible that the $(A_v, B_v, C_v, \sigma_v)$ parameters differ from fiducial D08 case, this has no impact on our conclusion that peaks number counts can discriminate among different MC relations.

\begin{figure*}
\resizebox{\hsize}{!}
{\includegraphics[width=7.0cm]{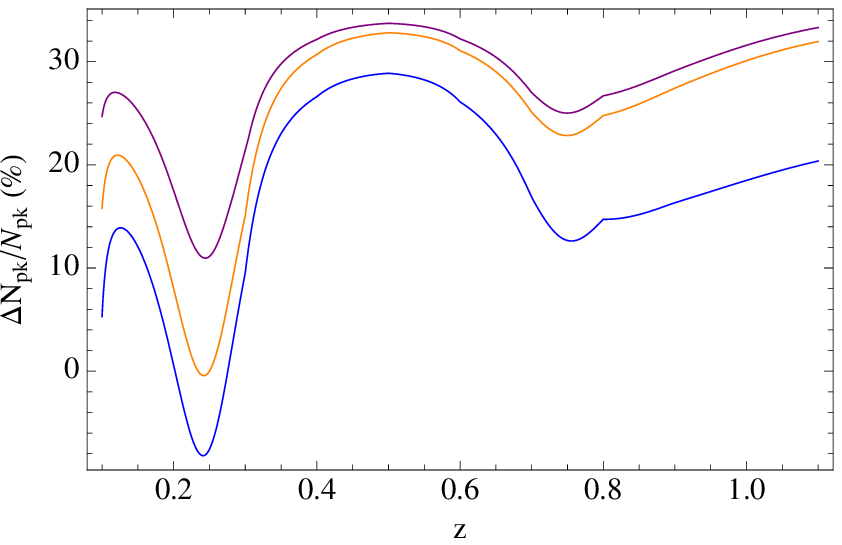}
\includegraphics[width=7.0cm]{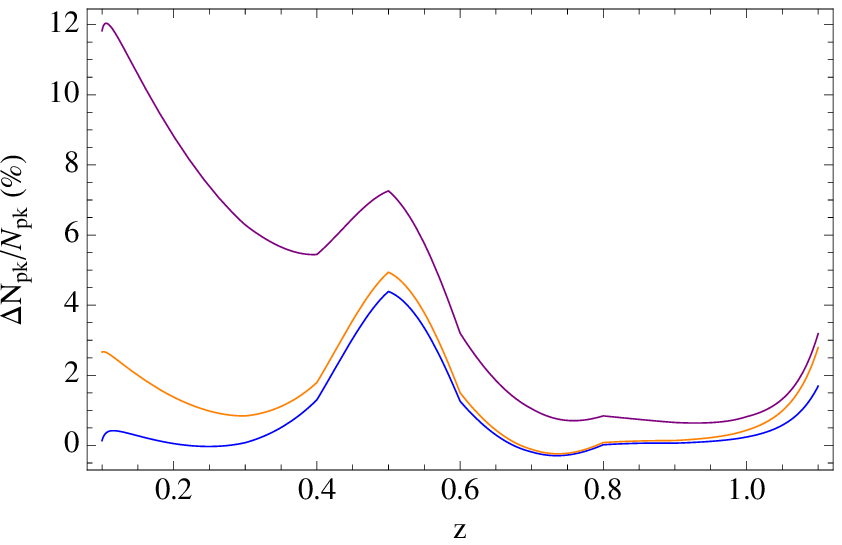}}
\caption{Percentage deviations $\Delta {\cal{N}}_{\rm pk}(z)/{\cal{N}}_{\rm pk}(z) = \left [ {\cal{N}}_{\rm pk}^{CDM}(z) - {\cal{N}}_{\rm pk}^{bar}(z) \right ]/{\cal{N}}_{\rm pk}^{CDM}(z)$ of the baryons included peak number counts from the CDM only case for the reference (blue), AGN80 (orange) and AGN85 (purple) models for peaks with $S/N \ge 3$ (left) or $S/N \ge 5$ (right).}
\label{fig: npkbar}
\end{figure*}

As discussed in Velliscig et al. (2014 and refs therein), baryons also change the mass function altering both the masses of single haloes and their abundances. Based on hydrodynamical simulations with different recipes for the details of baryons physics, Velliscig et al. (2014) provided approximated formul\ae \ to convert the CDM only mass function in its baryons included counterpart. Using their formalism, we have thus recomputed ${\cal{N}}_{\rm pk}(z)$ adopting the D08 MC relation and changing the baryon model.\footnote{We refer the reader to Velliscig et al. (2014) for the details of the baryons implementation for the three considered cases.} Fig.~\ref{fig: npkbar} shows $\Delta {\cal{N}}_{\rm pk}(z)/{\cal{N}}_{\rm pk}(z)$ taking the CDM only fiducial case as reference for two choices of the threshold $S/N$ value. While deviations from the CDM only case can be significant, they are nevertheless smaller than $12\%$ when only considering peaks with $S/N \ge 5$---which are the ones used in our Fisher matrix 
forecasts. 
We therefore expect that including baryons does not significantly change our results.

\subsection{The choice of the filter aperture and profile}
As a final remark, we want to qualitatively discuss how the choice of the filter impacts our results. Once the profile has been set (in our case resorting to the M05 optimal filter), one has only to choose the aperture $\vartheta$. Our choice of $\vartheta = 2 \ {\rm arcmin}$ motivated by the need to match the filter aperture to the typical scale radius $R_s$ of the NFW profile. Needless to say, $R_s$ is not the same for all clusters and, moreover, its value in angular units also depend on the cluster redshift and the background cosmology. Our choice is therefore based on a median cluster with mass $M_{200} \sim 5 \times 10^{14} \ {\rm M_{\odot}}$ at $z \sim 0.3$, where we expect the peak number counts to be the largest. Nevertheless, it is worth noting that this choice has not been done to maximise the $S/N$ ratio since this depend on the mass, the redshift and MC relation. What is important is that, no matter what the $\vartheta$ value is, the analysis presented here is still correct. Indeed, the Fisher 
matrix forecasts only rely on the assumption that the peaks detected are due to clusters and not fake ones. This is guaranteed by the choice of large enough $S/N$ threshold. If we had chosen a different $\vartheta$, we could have repeated the same analysis provided the threshold $S/N$ still selects only clusters peaks.

Although we do not aim here at optimising the filter aperture to minimise the constraints on the MC parameters, we nevertheless note that, holding fixed the threshold $S/N$ (which is a quite good assumption for $1 \le \vartheta/{\rm arcmin} \le 3$), the larger the aperture, the higher the total peak number. However, this does not automatically translate in stronger constraints. Indeed, changing the filter aperture also changes the orientation of the confidence contours in the projected two-dimensional parameter spaces. This result can be qualitatively explained by noting that, for a given redshift, the larger aperture, the smaller the minimum cluster mass to get a $S/N$ value larger than the threshold. A similar consideration also holds for the critical detection redshift at a given mass. As a consequence, the effective redshift of the peak sample and the mass regime investigated change with the aperture size, thus making the orientation of the contours different depending on the $\vartheta$ values. Since 
the magnitude of the constraints also depend on the mass and redshift regime explored, one cannot use a rule of thumb to conclude that larger apertures lead to smaller peak sample and weaker constraints.

The choice of the filter is even more subtle. Ideally, the optimal filter should guarantee both completeness and purity, that is to say it should make it possible to build up a catalogue containing all the peaks present in the survey with each one of them associated with a cluster and not due to large scale structure noise. Although the M05 optimal filter we use here was designed to fulfil both these criteria (the purity being guaranteed by the $S/N$ selection), it is nevertheless far from being perfect since it is based on the assumption that noise and signal combine in a linear way. Nonetheless, from our viewpoint, a less than ideal filter can still work to infer efficiently constraints on the parameters of interest. Indeed, what we need is a peak catalogue whose redshift distribution can be theoretically predicted without any unaccounted-for systematics. This is the case for the optimal filter in the high $S/N$ regime we have used here. Under these circumstances, the signal is dominated by the cluster 
itself so that it does not matter whether the noise combines linearly or not with it. As a result, the peak number counts are correctly predicted with the analytical formalism we use, so that we can confidently rely on the Fisher matrix forecasts we derive.

While the present paper was already completed, we came aware of a quite similar work by Mainini \& Romano (2014, hereafter MR14). They forecast the constraints on MC relation parameters from peak number counts in Euclid, but they used a different filter and different redshift scaling of the MC relation. Also, the background cosmological model and the $S/N$ threshold are different. As a result, a straightforward comparison is not possible. We nevertheless note that our constraints (for the case with MC and $(\Omega_M, \sigma_8)$ parameters left free to vary) are stronger than theirs. This is likely due to how the filter has been dealt with. In our approach, we assume that the filter is fixed to the one computed assuming a fiducial D08 MC relation so that $\hat{\Psi}(\ell)$ in Eq.~(\ref{eq: defsigmaratio}) is the same whatever MC relation is used to predict the peak number counts. Such a choice (mimicking what is actually done in building a catalogue from shear maps) allows to maximise the differences among MC 
relations. Since this were not done in MR14, their predicted peak redshift distribution is less dependent on the MC parameters---thus weakening the forecast constraints.

\section{Conclusions}
Weak lensing peak statistics in an Euclid\,-\,like survey will offer an effective tool to study the properties of galaxy clusters. In particular, the MC relation can be tightly constrained. Differently from ongoing studies which focus on the detailed study of a small sample of objects---usually of a few dozens of clusters---number counts can measure the MC relation in a statistical way by studying how it affects the number of thousands of detectable haloes and its evolution with redshift. The properties of the clusters are then measured from the whole population of haloes that pass the detectability threshold rather than from a limited sample of objects which might suffer from selection biases.

The massive end of the mass\,-\,concentration relation is of particular interest in the context of structure formation and evolution. The dynamical state of dark matter haloes might cause a non\,-\,monotonic relation between mass and concentration at high redshift \citep{2012MNRAS.423.3018P,2012MNRAS.427.1322L}. Massive systems at high redshift are likely still accreting material in a transient stage of high concentration before virialisation \citep{2012MNRAS.427.1322L}. Peak statistics will determine the MC slope to $\ls 0.03$ at $M_{200}>10^{15}M_\odot$ with clusters all over the redshift range, from $z\ls 0.1$ to $\gs 0.8$, contributing significantly to the total number of detectable peaks. This will then probe the assembly history of massive haloes.

Peak number counts are not only a probe of the MC relation, but also test the growth of structure through the dependence on the halo mass function. As we have shown, the peaks redshift distribution can nicely complement CMB data to further improve the constraints on the cosmological parameters. In particular, should the MC relation be known, peak number counts work better than the combination of current BAOs and SNeIa to constrain the present day value of the dark energy equation of state. Even in the least favourable case of fully unknown MC parameters to be marginalised over, peak number counts still nicely combine with Planck to reduce the error on dark energy parameters and further decrease those on the other cosmological parameters.

The constraints we have discussed have been obtained relying on the peak redshift distribution. However, this is only a zero\,-\,th order statistics, while higher order ones (such as the correlation function between peaks) are now being started to be developed \citep{2013MNRAS.432.1338M}. Although still in its infancy, such an approach is worth to be investigated in order to see how it depends on the halo properties and whether it can help to strengthen constraints on MC relation parameters. Furthermore, a combined analysis of cluster number counts from different tracers (weak lensing peaks, X\,-\,ray and Sunyaev\,-\,Zel'dovich) could represent a promising way to better constrain the scaling with mass and redshift of the MC relation, as each probe tests a different regime. In this case, one could finally find out which MC relation is the most reliable, hence opening the hunt for the physics needed to fill the gap between numerical results and the inferred MC relation.

\begin{acknowledgements}
VFC warmly thanks Silvia Galli for help with the Planck covariance matrix, Marco Velliscig for discussions on the impact of baryons, and Roberto Mainini for sharing his results in advance of publication. VFC is funded by Italian Space Agency (ASI) through contract Euclid\,-\,IC (I/031/10/0). VFC, RS and MS acknowledge financial contribution from the agreement ASI/INAF/I/023/12/0. SC is fundec by FCT\,-\,Portugal under Post\,-\,Doctoral Grant No. SFRH/BPD/80274/11. GC acknowledges support form the PRIN-INAF 2011 ``Galaxy Evolution with the VLT Survey Telescope (VST)''.
\end{acknowledgements}

\bibliographystyle{aa}
\bibliography{peaksmcaa2nd}

\end{document}